\documentclass[twocolumn,amsmath,amssymb]{revtex4-1}
\usepackage{graphicx}
\usepackage{dcolumn}
\usepackage{bm}

\begin{document}
\title{Hidden Variable Quantum Mechanics from Branching from Quantum Complexity}
\author{Don Weingarten}
\affiliation{donweingarten@hotmail.com}

\begin{abstract}
Beginning with the Everett-DeWitt many-worlds interpretation of quantum mechanics, there have
been a series of proposals for how the state vector of a quantum system might be split at any instant
into orthogonal branches, each of which exhibits approximately classical behavior.
Here we propose a decomposition of a state vector into branches 
by finding the minimum of a measure of the net quantum complexity of the branch 
decomposition.
We then propose a method for finding an ensemble of possible initial state vectors 
from which a randomly selected member, if evolved by ordinary Hamiltonian time evolution, 
will follow a single sequence of those branches of many-worlds quantum mechanics which 
persist through time. Macroscopic reality, we hypothesize, consists of an
accumulating sequence of such persistent branching results.
For any particular draw, the resulting deterministic system
appears to exhibit random behavior as a result of the successive emergence over time of information
present in the initial state but not previously observed. 
\end{abstract}

\maketitle
\section{\label{sec:intro}Introduction}

Microscopic particles have wave functions spread over all possible positions. 
Macroscopic objects simply have positions, or at least center-of-mass positions.
How to apply the mathematics of quantum mechanics to extract predictions 
registered in the macroscopic world of positions 
from experiments on microscopic systems having wave functions but not definite positions 
is well understood for all practical purposes.
But less well understood, or at least not a subject on which there is a clear consensus, is 
how in principle the definite positions of the macroscopic world emerge from the microscopic 
matter of which it is composed, which has only wave functions but not definite positions.
There is a long list of proposals.
In the present article we add another.  

We begin in Section \ref{sec:problems} with a brief reminder of ``the problem of measurement''
which arises for an experiment
in which a microscopic system interacts with a macroscopic measuring device 
with both systems assumed governed by quantum mechanics.
Among the proposals which address this problem
are the many-worlds interpretation \cite{Everett, DeWitt} and
environmentally-induced decoherence \cite{Zeh, Zurek, Wallace, Riedel}. 
Shared by these proposals is the hypothesis that the quantum state of the universe,
as time goes along, naturally splits into an set of orthogonal branch states each of which displays,
approximately, a distinct configuration of macroscopic reality.
We will argue, however, that rules according to which these proposals 
are to be applied to the world are intrinsically uncertain
and can be made precise only by the arbitrary choice of auxiliary parameters.
The uncertainty is not simply the approximate nature of the macroscopic
description of an underlying microscopic system, but rather that
the branching process of the microscopic system itself, in each of these proposals, occurs
according to uncertain rules.
And as a consequence, it seems to me
implausible that the corresponding branches are, by themselves, macroscopic reality.
In addition, missing from these proposals is a mathematical structure
that allows even the process of choosing the auxiliary parameters to be stated precisely.
These various limitations we will try to address in a sequence of several steps.

A main feature of the proposal we present here
is that branch formation does not follow from unitary time
evolution by itself nor does it 
entail a modification of unitary time evolution.
Instead, branch formation consists of an additional layer
of the world that sits on top of unitary time evolution.

In Section \ref{sec:complexity}, modifying ideas from \cite{Nielsen},
we define for a 1-dimensional lattice field theory of fermions
a version of quantum complexity
designed to measure the spatial structure of entanglement in a state vector.
In Section \ref{sec:entangledpoint} we define an entangled
state of a pair of particles each at a corresponding single point and then show in
Appendix \ref{app:entangledpoint}
that the state's complexity 
is bounded from below
and from above by quantities linear in the distance between the points.
In Section \ref{sec:entangledextended} we define an
entangled state of a pair of particles one of which has an
extended wave function and show
in Appendix \ref{app:entangledextended}
that the state's complexity 
is bounded from below and above by expressions 
linear both in the distance between particles and
in the width of the wave function.
In Section \ref{sec:branching} we then propose finding a branch decomposition of any state by 
minimizing a measure of the 
decomposition's net complexity.
The net complexity minimized depends on a parameter with units of length, the branching threshold $b$, which
determines the circumstances under which quantum behavior crosses over to classical. 
The branching threshold $b$ should in principle be measurable.

In Section \ref{sec:dqm} using the decomposition of Section \ref{sec:complexity},
we propose a branch decomposition at late time $t_{out}$, restored to some finite time
$t_{in}$ by unitary time evolution, then subjected to the limit $t_{out} \rightarrow \infty$,
thereby selecting only those branching results which persist through time.
The set of branches found directly at a fixed finite time, we will argue,
in some cases is not covariant, a
problem potentially avoided by the late time limit.
Macroscopic reality, we then conjecture, consists of the
accumulating sequence over time of such permanent branch results.
Each of the vectors in the limiting decomposition at $t_{in}$, according to this hypothesis,
if evolved by the usual unitary time evolution  
will follow only a single sequence of persistent branch results \cite{Stoica}.
Macroscopic reality still emerges as an approximate description of
a corresponding state vector's time evolution. But the underlying state
vector itself evolves purely deterministically through a time trajectory of branching
events without additional approximations or missing parameters.

In Section \ref{sec:stern} 
we consider the theory of Section \ref{sec:dqm} applied to a simplified Stern-Gerlach experiment.
In Section \ref{sec:bell} we apply the theory of Section \ref{sec:dqm} 
to a model of a test of Bell's theorem 
and obtain
the expected quantum mechanical results.
We conclude in Section \ref{sec:conclusion} with comments
about issues yet to be addressed.

Throughout this paper, we will assume flat space-time. In Section \ref{sec:dqm}
the limit $t_{out} \rightarrow \infty$ ignores events on a cosmological time scale.

\section{\label{sec:problems}Problems}
Let $\mathcal{S}$ be a microscopic system to be measured, with corresponding state space $\mathcal{H}_\mathcal{S}$,
for which a basis is $\{|s_i>\}, i > 0$.
Let $\mathcal{M}$ be a macroscopic measuring device with corresponding state space $\mathcal{H}_\mathcal{M}$
containing the set of vectors $\{ |m_i>\}, i \geq 0$.  For each different value of $i > 0$ the state $|m_i>$ is 
a macroscopically distinct meter reading.
Let $|m_0>$ be an initial state showing no reading. In the 
combined system-meter product state space 
$\mathcal{H}_\mathcal{S} \otimes \mathcal{H}_\mathcal{M}$, a 
measurement of $\mathcal{S}$ by $\mathcal{M}$ over some time interval
takes each possible initial state $|s_i> |m_0>$ into the corresponding final state $|s_i>|m_i>$
with the measuring device displaying the measured value of the microscopic system's variable
\begin{equation}
|s_i> |m_0> \rightarrow |s_i> |m_i>.
\end{equation}
By linearity of quantum mechanical time evolution, however, it then follows that a measurement
with a linear superposition in the initial state will yield a final state also with a superposition
\begin{multline}
[\alpha|s_1> + \beta|s_2>] |m_0> \rightarrow \\ 
\alpha |s_1> |m_1> + \beta|s_2> |m_2>. 
\end{multline}
In the measured final state, the meter no longer has a single value but a
combination of two values which cannot, by itself,  be connected to a recognizable configuration of a macroscopic object.
The absence of a recognizable configuration for the macroscopic device is the ``problem of measurement''.

The resolution of this problem proposed by the many-worlds interpretation of quantum mechanics \cite{ Everett, DeWitt} 
is that the states $|s_1> |m_1>$ and $|s_2> |m_2>$ actually represent two different worlds. In each world the 
meter has a definite position but with different positions in the two different worlds. For an interaction
between two systems, the splitting into separate worlds is done in the Schmidt basis, in which the density 
matrix of the measured system is diagonalized.  Among the problems of the many-worlds interpretation, however,
is that in general, for plausible models of a measurement process, 
the individual worlds given by the Schmidt basis do not have sufficiently narrow coordinate dispersions to count as classical reality \cite{Page}.  In addition, it is unclear under
what circumstances and according to what basis a system larger than just a micro system and a measuring 
device should be split into separate worlds.

A resolution to the first of these problems, the absence of 
classical behavior in the split branches, is proposed to occur through environmentally-induced decoherence 
\cite{Zeh, Zurek}.
According to this proposal, the system-meter combination should not be considered in isolation
but instead an account is required of the rest of the macroscopic environment with which the meter can interact.
When the value of a macroscopic meter is changed by recording the value of a microscopic coordinate, the meter's new state
rapidly becomes entangled with a large number of other degrees of freedom in the environment 
\begin{multline}
\label{entangled}
[\alpha |s_1> |m_1> + \beta |s_2> |m_2>] |e_0> \rightarrow \\
\alpha|s_1> |m_1> |e_1> +  \beta |s_2> |m_2> |e_2>. 
\end{multline}
For a particular choice of bases for system, meter and environment, determined by the combined system's dynamics, entanglement
of the meter with the environment proceeds as quickly as possible,
$| e_1>$ and $| e_2>$ almost do not mix in the course of further time development,
and $|e_1>$ and $|e_2>$ include many redundant copies
of the information in $|s_1> |m_1>$ and $|s_2> |m_2>$, respectively.
Based on these various considerations it is argued that entangled environmental states
$|e_1>$ and $|e_2>$
behave essentially as permanent classical records of the experimental results.
Correspondingly, for many-worlds augmented with decoherence \cite{Wallace}, the circumstance 
under which a system splits into distinct worlds is when a superposition
has been produced mixing distinct values of one of these effectively classical 
degrees of freedom. Each distinct value of the coordinate in such a superposition
goes off into a distinct world. 

A step toward resolving the second problem, the absence of a criterion for branching
for the universe as a whole rather than simply for some system-apparatus pair, takes the form
of a theorem \cite{Riedel} according to which, for a system as a whole, 
if a particular spatial pattern of redundant records happens to occur,
then there is a unique corresponding decomposition of a state vector into effectively
classical branches.

A residual problem of \cite{Zeh, Zurek, Wallace, Riedel}, however, 
is that the rules governing their application to the world are intrinsically uncertain.
In particular, the record production needed for 
environmentally-induced decoherence occurs over some nonzero intervals of time
and space, and perhaps is entirely completed only asymptotically in 
long time and large distance limits.
What fraction of the initial state in Eq. (\ref{entangled})
must become entangled with the environment for splitting into
classical branches to occur?
When exactly over the time interval of decoherence, does the splitting of the world in parts 
occur? And since the process extends over space, this timing will differ in different
Lorentz frames. Which is the correct choice? 
These various questions may be
of no practical consequence in treating the meter readings as nearly classical
degrees of freedom after entanglement and using the resulting values to formulate observable predictions.
But what is clear is that something is missing from the theory. From outside the theory,
something additional and arbitrary needs to be supplied by hand to resolve each of these issues.
Moreover, no mathematical machinery is present in any of these proposal
which allows the process of filling in what is missing to be stated precisely.
As a consequence of all of which it seems to me implausible that the branches
provided by these accounts are by themselves the fundamental substance of reality.

A discussion of issues concerning environmentally induced decoherence
and its combination with the many-worlds interpretation of quantum mechanics
appears in \cite{Schlosshauer}.

The remainder of this paper consists mainly 
of setting up the machinery for two conjectures intended to offer a way
to fill in the omissions which we believe
prevent environmentally induced decoherence in its present
formulation from providing an account of macroscopic reality.
The two conjectures appear in Section \ref{subsec:branchingatinfinity}.

\section{\label{sec:complexity}Complexity}

We now construct a complexity measure
for a 1-dimensional lattice field theory of fermions, at a single instant
of time, modifying ideas from \cite{Nielsen}.
No feature of the discussion depends significantly on the lattice dimension or on the choice of fermions.
Complexity measures similar to the one we will discuss can
also be constructed in higher dimensions and for theories including bosons.

\subsection{\label{subsec:hilbertspace} Lattice Fermion Hilbert Space}

The state space for the system $\mathcal{H}$ we assume is
a tensor product of spaces $\mathcal{H}_x$ one for each integer valued lattice point $x$ in some
arbitrarily large but finite range $|x| \le x_{max}$
\begin{equation}\label{tensorproduct}
\mathcal{H} = \otimes_x \mathcal{H}_x.
\end{equation}
An orthogonal basis for each $\mathcal{H}_x$ consists of the
vacuum $|0>_x$, 1-particle spin up and spin down states
$|1>_x$ and $|-1>_x$, and a 2-particle state $|2>_x$, with one spin up and the other down.
The overall system vacuum $| 0>$ is
$\otimes_x |0>_x$. Let $N_x$ be the operator on $\mathcal{H}$ giving the fermion
count at site $x$ and let $N$ be the total fermion count
\begin{subequations}
\begin{eqnarray}
\label{numberx} 
\hat{N}_x & = & N_x \otimes_{q \neq x} I_q, \\
\label{numbertot}
N & = & \sum_x \hat{N}_x.
\end{eqnarray}
\end{subequations}
For convenience we now write both $\hat{N}_x$ and $N_x$ as $N_x$.

We define in $\mathcal{H}$ a set $\mathcal{P}$ of product states.
Define the operators $a( x, i)$ for $i$ of -1 or
1 which annihilate the vacuum $|0>$,
their adjoints, $a^\dagger( x, i)$, which create single particle
states and together have the 
anticommutation relations
\begin{subequations}
\begin{eqnarray}
\label{antiaa}
\{ a( x, i), a(y, j) \} & = & 0, \\
\label{antiadad}
\{ a^\dagger( x, i), a^\dagger(y, j) \} & = & 0, \\
\label{antiada}
\{ a^\dagger( x, i), a(y, j) \} & = & \delta_{x y} \delta_{i j}.
\end{eqnarray}
\end{subequations}
From these it is convenient to define $b^\dagger( x, i)$ to be the
same as $a^\dagger( x, i)$ for $i$ of 1 or -1 and for $i$ of 2
\begin{equation}
\label{b2}
b^\dagger( x, 2) = a^\dagger( x, -1) a^\dagger( x, 1).
\end{equation}
For a
product in order of increasing $x_j$ with $i_j \ne 0$ we then adopt
the sign convention
\begin{equation}
\label{defadagger}
\prod_j b^\dagger( x_j, i_j) | 0> = \otimes_j |i_j>_{x_j} \otimes_{q \ne x_0, .. x_n} | 0>_q.
\end{equation}

For a normalized position wave function $p(x)$, 
and a normalized spin wave function $[ s(1), s(-1) ]$,  define creation operators 
$c^\dagger( p, s)$ for extended states
\begin{equation}
\label{extended}
c^\dagger( p, s) = \sum_{x i} p(x) s(i) a^\dagger( x, i).
\end{equation}
From a sequence of n position and spin wave functions, define an n-particle product state to be
\begin{equation}
\label{productstate}
c^\dagger( p_{n-1}, s_{n-1}) ... c^\dagger( p_0, s_0) |0>.
\end{equation}
Let $\mathcal{P}$ be the set of all product states. 

\subsection{\label{subsec:operatorspace} Hermitian Operator Hilbert Space}

We now define a Hilbert space over the reals of Hermitian operators
acting on $\mathcal{H}$. 
For any site $x$, let
$\mathcal{F}_x$ be the set of traceless Hermitian operators acting on
$\mathcal{H}_x$ which commute with $N_x$.
For any pair of nearest neighbor sites $x < y$, let
$\mathcal{F}_{xy}$ be the set of traceless Hermitian operators acting on
$\mathcal{H}_x \otimes \mathcal{H}_y$ which commute with $N_x + N_y$.
$\mathcal{F}_{xy}$ can be made into a Hilbert space over the reals with the inner product
and norm
\begin{subequations}
\begin{eqnarray}
\label{deffg}
< f, g> & = & Tr( f g), \\
\label{defnorm}
\parallel f \parallel ^2 & = & < f, f>.
\end{eqnarray}
\end{subequations}
Any $f \in \mathcal{F}_x$ can be viewed as $f \otimes I_y \in \mathcal{F}_{xy}$,
and any $f \in \mathcal{F}_y$ can be viewed as $I_x \otimes f \in \mathcal{F}_{xy}$.
The norm in Eq. (\ref{defnorm}) of $f \otimes I_y \in \mathcal{F}_{xy}$
for the nearest neighbor pair $x < y$
is the same as the norm of $I_y \otimes f \in \mathcal{F}_{yx}$ for the
nearest neighbor pair $y < x$.
Define $\mathcal{G}_{xy}$ to be the subspace of 
$\mathcal{F}_{xy}$
orthogonal to $\mathcal{F}_{x}$, and $\mathcal{F}_{y}$.

The operators and inner product just defined can be extended to the full space $\mathcal{H}$.
For any $f_x$ in some $\mathcal{F}_x$ and any $g_{xy}$ in some $\mathcal{G}_{xy}$, define
corresponding operators on $\mathcal{H}$ by
\begin{subequations}
\begin{eqnarray}
\label{defhf}
\hat{ f}_x & = & f_x \otimes_{q \ne x} I_q, \\
\label{defhg}
\hat{ g}_{x y} & = & g_{x y} \otimes_{q \ne x, y} I_q.
\end{eqnarray}
\end{subequations}
Let $K$ be a vector space over the reals of traceless, Hermitian linear
operators $k$ on $\mathcal{H}$ commuting with $N$ given by sums
of the form
\begin{equation}
\label{defk}
k = \sum_x \hat{f}_x + \sum_{x y} \hat{g}_{x y},
\end{equation}
for any collection of $f_x \in \mathcal{F}_x$ on a set of distinct sites $\{x\}$,
and any collection of 
$g_{x y} \in \mathcal{G}_{x y}$ on a set of distinct nearest neighbor pairs $\{(x, y)\}$.
Pairs in the set $\{(x, y)\}$ may overlap on a single site but not on both.
The inner product and norm of Eqs. (\ref{deffg}) and 
(\ref{defnorm}) can then be extended to $k$ of Eq. (\ref{defk})
and $k'$ defined similarly according to
\begin{subequations}
\begin{eqnarray}
\label{defkkprime}
< k, k'> & = & \sum_x < f_x, f'_x> + \sum_{x y} < g_{x y}, g'_{x y}>, \\
\label{defnormk}
\parallel k \parallel^2 & = & \sum_x < f_x, f_x> + \sum_{x y} < g_{x y}, g_{x y}>.
\end{eqnarray}
\end{subequations}

The inner product of Eq. (\ref{defkkprime})
is equivalent to
\begin{equation}
\label{defkkprime1}
< k, k'> = 4^{1 - 2 x_{max}} Tr( k k'),
\end{equation}
for $k$ and $k'$ in Eq. (\ref{defk}), the trace in this case taken over all of
$\mathcal{H}$. The equality of
the inner products of Eqs. (\ref{defkkprime}) and (\ref{defkkprime1}) follows
from the requirement that all $\mathcal{F}_x$ and $\mathcal{G}_{x y}$ are traceless,
that all $\mathcal{G}_{x y}$ are orthogonal to the corresponding 
$\mathcal{F}_x$ and $\mathcal{F}_y$ and that the trace
of the identity operator $I_x$ on any $\mathcal{H}_x$ is 4.

Although Eq. (\ref{defkkprime1}) could be
taken as an alternative definition of $<k, k'>$ for
the space $K$ of operators on the fermion Hilbert space
of Section \ref{subsec:hilbertspace}, 
for a corresponding space of operators
on a boson Hilbert space, the underlying $\mathcal{H}_x$
are infinite dimensional, the trace of the identity operator 
$I_x$ on any $\mathcal{H}_x$ is not finite and
there is no equivalent of Eq. (\ref{defkkprime1}).
We thus prefer to view Eqs. (\ref{defkkprime}) and (\ref{deffg}) as
the definition of $<k, k'>$.

\subsection{\label{subsec:complexitydef} Complexity from Unitary Trajectories}

From this machinery, for any pair of states $| \psi_0>, |\psi_1> \in \mathcal{H}$ with equal
particle number we define 
the complexity $C(|\psi_1>, |\psi_0>)$ of $|\psi_1>$ with
respect to $|\psi_0>$. 
For $0 \leq t \leq 1$, let $k( t) \in K$ be a piecewise continuous trajectory of operators.
Let the unitary operator $U_k(t)$ on $\mathcal{H}$ be the solution to the time development
equation and boundary condition
\begin{subequations}
\begin{eqnarray}
\label{udot}
\dot{U}_k(t) & = &-i k( t) U_k( t), \\
\label{uboundary0}
U_k( 0) & = & I.
\end{eqnarray}
\end{subequations}
Among such $k(t)$, we consider only those which in addition fulfill
\begin{equation}
\label{uboundary1}
\xi U_k( 1) |\psi_0> = |\psi_1>,
\end{equation}
for some complex number $\xi$ with $| \xi | = 1$.
Then the complexity $C(|\psi_1>, |\psi_0>)$ is the minimum over all such
$k(t)$ of the integral
\begin{equation}\label{complexity}
C(| \psi_1>, |\psi_0>) = \min \int_0^1 d t \parallel k( t) \parallel. 
\end{equation}

Finally, any product state in $\mathcal{P}$ we assign 0 complexity. 
The complexity $C( |\psi_1>)$ of any state $|\psi>$ not in $\mathcal{P}$
is defined to be the distance to the closest product state
\begin{equation}
\label{cpsi1}
C( |\psi_1>) = \min_{|\psi_0> \in \mathcal{P}} C(| \psi_1>, |\psi_0>).
\end{equation}

For any pair of vectors $|\psi_1>, |\psi_0> \in \mathcal{H}$ with equal
particle numbers, an operator trajectory $k(t)$
giving $|\psi_1> = \xi U_k(1)|\psi_0>$ can be
found even though $K$ in Eq. (\ref{defk}) is restricted
to nearest neighbor Hamiltonians. We show in Appendix \ref{app:complexitygroup}
that the group $G$ of all $U_k( 1)$ realizable as solutions to Eqs. (\ref{udot}) and
(\ref{uboundary0}) has as a subgroup the direct product
\begin{equation}
\label{formofg}
\hat{G} = \times_n SU(d_n),
\end{equation}
where $SU(d_n)$ acts on the subspace of $\mathcal{H}$
with eigenvalue $n$ of the total number operator $N$,
$d_n$ is the dimension of this subspace, and the product
is over the range $ 1 \le n \le 4 * x_{max} + 1$.
Missing from this range of $n$ are only two states, 
the vacuum, $n = 0$, and the state with every site occupied by
two particles, $n = 4 * x_{max} + 2$.

The restriction in Eq. (\ref{defk}) to nearest neighbor Hamiltonians
which preserve particle number
is a departure from \cite{Nielsen}.
What is gained from the combination of these restrictions
is that, as a consequence,
state vectors $|\psi>$ which carry entanglement over long distances
require $k(t)$ with many steps and thus
are assigned high complexity.  
In Sections \ref{sec:entangledpoint} and \ref{sec:entangledextended}
we consider two examples of states composed of a sum of a pair of 
product states, each of which individually has zero complexity, but for 
which the sum exhibits entanglement and as a result has
complexity which is bounded both from below and
from above by functions which grow linearly with the size of the region over
which the entanglement extends.

\section{\label{sec:entangledpoint} Complexity of Entangled Particles at Distant Points}

Let the entangled 2 particle state $|\omega>$ be
\begin{subequations}
\begin{eqnarray}
\label{omega0}
|\omega_0> & = & |-1>_0 \otimes_{q \ne 0, n} | 0>_q \otimes |-1>_n, \\
\label{omega1}
|\omega_1> & = & |1>_0 \otimes_{q \ne 0, n} | 0>_q \otimes |1>_n, \\
\label{omega}
|\omega> & = & \frac{1}{\sqrt{2}} | \omega_0> + \frac{1}{\sqrt{2}} |\omega_1>,
\end{eqnarray}
\end{subequations}
for some $n \ge 2$.
We show in Appendix \ref{app:entangledpoint} 
\begin{equation}
\label{cbound1}
C( |\omega>) \ge \frac{ n \pi}{ 8 \sqrt{2}}.
\end{equation}
The complexity of $|\omega>$ grows at least linearly with the distance between
the entangled pair of particles.

We obtain also in Appendix \ref{app:entangledpoint} 
an upper bound on $C(|\omega>)$ linear in the distance between the
pairs
\begin{equation}
\label{cbound2}
C( |\omega>) \le (n - 1 + \frac{1}{2\sqrt{2}}) \pi.
\end{equation}

\section{\label{sec:entangledextended} Complexity of an Entangled Extended State}

Although product states with extended wave functions are defined to have 0 complexity independent of
the width the wave function, 
for entangled states with extended wave functions for one or both of the particles there is a complexity cost
determined both by the distance between the states and by the width of the wave functions.

For simplicity, we present only the case with one of the two entangled particles in an extended state. 

Consider a version of 
Eqs. (\ref{omega0}) - (\ref{omega}) with one of the particles in an extended state
\begin{subequations}
\begin{multline}
\label{omegaprime0}
|\omega'_0>  = | -1>_0 \\ \otimes ( \frac{1}{\sqrt{r}} \sum_{ n \le x < n + r}|-1>_x \otimes_{ q \ne 0, x} | 0>_q),
\end{multline}
\begin{multline}
\label{omegaprime1}
|\omega'_1> = | 1>_0 \\ \otimes ( \frac{1}{\sqrt{r}} \sum_{ n \le x < n + r}|1>_x \otimes_{ q \ne 0, x} | 0>_q),
\end{multline}
\begin{multline}
\label{omegaprime}
|\omega'>  = \frac{1}{\sqrt{2}} | \omega'_0> + \frac{1}{\sqrt{2}} | \omega'_1>.\hphantom{dddddddddd}
\end{multline}
\end{subequations}

In Appendix \ref{app:entangledextended} we prove
\begin{equation}
\label{cbound1extended}
C( |\omega'>) \ge \frac{n \pi}{8 \sqrt{2}} + \frac{r \kappa}{2 \sqrt{2}}.
\end{equation}
where $\kappa$ is
\begin{equation}
\label{defkappa}
\kappa = \frac{1}{r} \sum_{0 < s < r} \arcsin( \sqrt{\frac{s}{2r}}).
\end{equation}
For large $r$, $\kappa$ approaches
\begin{equation}
\label{defkappalim}
\lim_{r \rightarrow \infty} \kappa = \int_{0}^{1} \arcsin( \sqrt{\frac{x}{2}}) d x.
\end{equation}

The complexity of an entangled pair with one of the particles in an
extended state grows at least linearly both with the distance between
the pairs and with the width of the extended state.

In Appendix \ref{app:entangledextended} we prove also
\begin{equation}
\label{cupperprime}
C( |\omega'>) \le ( n - 1 + \frac{\pi}{2 \sqrt{2}}) \pi + 2 \lambda r,
\end{equation}
where
\begin{equation}
\label{defkappaprime}
\lambda = \frac{1}{r} \sum_{0 < s < r} \arcsin(\sqrt{ \frac{s}{s +1}}).
\end{equation}
For large $r$ for most of the range $0 < s < r$, the argument of the sum in Eq. (\ref{defkappaprime}) is nearly $\frac{\pi}{2}$.
Therefore
\begin{equation}
\label{limkappaprime}
\lim_{r \rightarrow \infty} \lambda = \frac{\pi}{2}.
\end{equation}

\section{\label{sec:branching}Branching}

Using the complexity measure of Section \ref{sec:complexity} we now propose a
decomposition of a state vector into approximately classical branches. 

\subsection{\label{subsec:branchcomplexity} Net Complexity of a Branch Decomposition}

For any $|\psi> \in \mathcal{H}$ let 
 $ |\psi> = \sum_i |\psi_i>$
be a candidate orthogonal decomposition into branches.
Let $Q( \{|\psi_i>\})$ be a measure of complexity of this decomposition
\begin{multline}\label{defQ} 
Q( \{|\psi_i>\})  =  \sum_i < \psi_i | \psi_i> C( |\psi_i>) - \\
 b \sum_i < \psi_i | \psi_i> \ln( < \psi_i |\psi_i>),
\end{multline} 
with branching threshold $b > 0$. For any choice of $b$, the branching
decomposition of $|\psi>$ is defined to be the $\{|\psi_i> \}$ which minimizes
$Q(\{|\psi_i> \})$. The first term in Eq. (\ref{defQ}) is the mean complexity
of the branches split off from $|\psi>$. But each branch can also be thought
of as specifying, approximately, some macroscopic classical configuration of the
world. The second term represents the entropy of this random ensemble
of classical configurations.

Since $\mathcal{H}$ in Section \ref{subsec:hilbertspace} is finite 
dimensional,
the search to minimize $Q(\{|\psi_i>\})$ is over a compact set of possible
orthogonal decompositions. Since $Q( \{ |\psi_i> \})$ is nonnegative,
it must have at least one minimum on this set. We will assume without proof
that this minimum is unique except for a set of $|\psi>$ of measure 0.

For $b$ either 
extremely small or extremely large, the branches which follow from Eq. (\ref{defQ}) 
will
look nothing like the macro reality we see.  For small enough $b$,
the minimum of $Q( \{|\psi_i>\})$ will be dominated by the complexity term.
It follows from the discussion of Section \ref{sec:complexity}
that the minimum of the complexity term will occur for a set of branches each of which is nearly
a pure, unentangled multi-particle product state. Thus bound states
will be sliced up into unrecognizable fragments. On the
other hand, for very large $b$, the minimum of $Q( \{|\psi_i>\})$
will be dominated by the entropy term, leading to only 
a single branch consisting of the entire coherent quantum state. 
Again, unlike the world we see.

The result of all of which is that for the branches given by minimizing $Q( \{|\psi_i>\})$
of Eq. (\ref{defQ}) to 
have any chance of matching the macro world, $b$ has to be some finite
number. 

\subsection{\label{subsec:timeevolution} Time Evolution of Optimal Branch Decomposition}

Suppose $Q(\{|\psi_i>\})$ is minimized at each $t$ for
some evolving $|\psi(t)>$.
Given that time evolution is continuous, 
the optimal branch configuration will
be a piecewise continuous function of time. For an interval
the optimal configuration will move continuously, then at some isolated
point in time a configuration far from the optimal configuration
just preceding that point in time will acquire a lower
value of $Q(\{|\psi_i>\})$. The optimal 
configuration will then jump discontinuously
to the new configuration.

Consider the circumstance
under which the minimization of $Q( \{|\psi_i>\})$ would
cause some branch $|\psi_i>$ to split into
two pieces
\begin{equation}\label{splitpsi}
|\psi_i> = |\psi_i^0> + |\psi_i^1>.
\end{equation}
The terms in $Q( \{|\psi_i>\})$ arising from $|\psi_i>$ before
the split are
\begin{equation}\label{beforesplit}
<\psi_i|\psi_i>[ C( |\psi_i>) - b \ln( < \psi_i | \psi_i>].
\end{equation}
The terms from $|\psi_i^0>$, $|\psi_i^1>$ after the split can be written in the form
\begin{multline}\label{aftersplit}
<\psi_i|\psi_i>[ \rho C( |\psi_i^0>) + ( 1 - \rho) C( |\psi_i^1>) - \\
b \rho \ln( \rho) - b ( 1 - \rho) \ln( 1 - \rho)  - b \ln( < \psi_i| \psi_i>],
\end{multline}
where it is convenient to express $< \psi_i^0 | \psi_i^0>$ as $\rho < \psi_i | \psi_i>$.
Thus a split will occur if
\begin{multline}\label{splitcondition}
C( |\psi_i>) - \rho C( |\psi_i^0>) - ( 1 - \rho) C( |\psi_i^1>) > \\
-b \rho \ln( \rho) - b ( 1 - \rho) \ln( 1 - \rho).
\end{multline}
The condition for a split is a saving in complexity
by an amount linear in $b$. 
Splitting occurs as soon as a certain threshold amount
of complexity can be saved by the split. 
The largest possible value of
this threshold is for $\rho$ of $\frac{1}{2}$, in which case
it is given by $\ln(2) b$. 
Correspondingly no 2-way split will ever be accompanied by a complexity drop
of more than $\ln(2) b$. Averaged over $\rho$, the complexity drop
required by Eq. (\ref{splitcondition}) is $\frac{b}{2}$.

A split can also reverse itself if as a result of time evolution
the complexity of some branch changes sufficient to
reverse the inequality in Eq. (\ref{splitcondition}). 
It is also possible in principle for a set of some n branches
jointly to rearrange into a new configuration of n+1 branches for an
accompanying saving of up to $\ln(n + 1) b$. 
Both of these complications will be removed 
in Section \ref{sec:dqm} as a by product of
the hypothesis that
branch formation minimizing $Q(\{|\psi_i>\})$ 
occurs only at a single asymptotically late time.
This hypothesis we introduce primarily 
to try to avoid a potential 
problem with Lorentz covariance
of branching at any fixed finite time.

The hypothesis that branching occurs only at a single
asymptotically late time, however, adds a
complication to measurement of $b$ by observation
at finite time. We will briefly return to this
subject in Section \ref{sec:conclusion}.

\section{\label{sec:dqm}Deterministic Quantum Mechanics}

So far, we have defined complexity and branching only 
for a lattice field theory. But given
that the definition relies only of a field theory at a single instant of time 
and is thus independent of the system's dynamics, it seems
a plausible guess that the continuum limit of this construction 
should exit.
The subject of the continuum limit of
$Q( \{ |\psi_i> \})$ we will return to elsewhere.  

\subsection{\label{subsec:lorentzcovariance} Lorentz Covariance Problem for Branching}

Assuming this goes through without a hitch,
however,
there is still a problem with the Lorentz transformation
behavior of the branch construction rule. A discussion of this issue in
a slightly different setting and a solution somewhat related to
the one we will propose appear in \cite{Kent}.

The problem is actually a version of the original EPR paradox.
In two different Lorentz frames, starting from
initial states vectors related by a Lorentz transform, 
assume Hamiltonian time evolution and, in each frame,
assume branching which minimizes the net complexity
function $Q(\{|\psi_i>\})$ in Eq. (\ref{defQ}).
For some period of time assume the branching
in each frame is also related by a Lorentz transform.
But then at a pair of points separated
by a spacelike interval, suppose processes occur each of which,
by itself, is sufficient to cause splitting of a branch
the two events share. Assume in addition, that in one
frame one of these events occurs first but, 
in the other frame, as a consequence of the Lorentz
transform of their spacelike separation, the other event occurs first.
The result will be that in the time interval between the
events, the branch structure seen by the two different Lorentz
frames will be different. But our goal is to be able
to interpret branch state vectors as the underlying 
substance of reality. That interpretation clearly
is not possible if branch structure is different
according to different Lorentz observers.

This variant of the EPR paradox, in only slightly different clothing, 
we already briefly mentioned in Section \ref{sec:problems} and is a general problem for any formulation 
of branches as the substance of reality \cite{Zeh, Zurek, Wallace, Riedel} and not
only a problem for branches defined by minimization 
of $Q( \{ |\psi_i> \})$.

\subsection{\label{subsec:branchingatinfinity} Infinite Time Limit of Branching}

What to do? A possible approach is suggested by the observation
that for any distinct pair of Lorentz frames, for some $t_{out}$ sufficiently
late, all interactions which cause permanent branch splitting will have finished in both frames. 
Thus, at least for that pair of frames, the difficulty
in Section \ref{subsec:lorentzcovariance} might potentially be avoided.

In any frame, for $t_{out}$ sufficiently late, the corresponding state $|\psi_{out}>$ 
will consist of individual particles and bound systems, possibly
including some collection of chunks of solids, spread
out over large distances, all moving away from each other.
Whatever entanglement may have arisen from interactions, however,
will remain recorded in the state's structure. An example
of this will be considered in Section \ref{sec:stern}.

As a first step toward Lorentz covariance, we now add to
Eq. (\ref{defQ}) the requirement that for each $|\psi_i>$,
the complexity $C(|\psi_i>)$ be evaluated in the 
rest frame of $|\psi_i>$.

We now minimize $Q( \{ |\psi_{out, i} \})$ 
to find a corresponding branch
decomposition $\{ |\psi_{out, i}> \}$ of $| \psi_{out}>$, 
translate the result back to some some fixed $t_{in}$ to define
a decomposition $\{ |\psi_{in, i}> \}$ of $| \psi_{in}>$
\begin{equation}\label{inbranches}
|\psi_{in, i}> = \exp[ i ( t_{out} - t_{in}) H ] |\psi_{out, i}>,
\end{equation}
and take the limit of $t_{out} \rightarrow \infty$.
This limit is needed to insure
that $t_{out}$ is after the last persistent branching event
for any state in any frame.
It is also required to
avoid some small but nonzero dependence of the
$\{ |\psi_{in, i}> \}$ ensemble on the choice of $t_{out}$.
Any branches which
may have formed but then recombined will not be included in 
$\{ |\psi_{out, i}> \}$. Only those which survived will appear.

Our conjecture, the
first of the two
mentioned in Section \ref{sec:problems},
is that for a
pair of frames related by
some Lorentz transform $L$,
branches in the set 
$\{ |\psi_{in,i}> \}$ found
in one frame and in the set 
found in the other frame $\{ |\psi'_{in,i}> \}$
can be arranged in corresponding pairs
related a unitary operator $U( L)$
determined by $L$
\begin{equation}
\label{branchrelation}
|\psi'_{in, i}> = U( L) |\psi_{in, i}>.
\end{equation} 

In any particular Lorentz frame the state of the real world at $t_{in}$, 
we now take as a random draw from the ensemble
$\{ |\psi_{in, i}> \}$, each weighted 
with the corresponding probability $ <\psi_{in, i}|\psi_{in, i}>$.
For any time other than $t_{in}$, the state
chosen is evolved continuously in time with no further
branching
\begin{equation}\label{evolution}
|\psi_i(t) > = \exp[ -i ( t - t_{in}) H ] |\psi_{in, i}>.
\end{equation}

The branches of a state which form at some event can be thought of
as recording different values of a corresponding meter.
An accumulating sequence of persistent branching results
can therefore be viewed also as an accumulating sequence of persistent records.
We  now introduce the second of the two conjectures mentioned
in Section \ref{sec:problems}, which is that
possible histories of macroscopic reality are
accumulating sequences of those branching results
which persist through time. Each
final branch $|\psi_{out, i}>$ will record one such
macroscopic history.
Each corresponding $|\psi_{in, i}>$, evolved through time, will then
reproduce a corresponding single history of macroscopic branch choices.

In a world which obeys purely classical mechanics, the configuration of 
the world on any time slice can be used to determine the configuration
on all past time slices and, in effect, is a persistent
record of all past configurations.
The hypothesis that the 
macroscopic part of the quantum mechanical
world is the accumulating set of persistent branches can be viewed as
a kind of extension of what holds in a purely classical world.

As mentioned in passing in 
Section \ref{sec:problems}, 
the formation of
permanent records is also an identifier of
quantum degrees of freedom which become classical degrees of freedom
according to environmentally 
induced decoherence in
\cite{Zurek}.

Each final state, at some late time
after all branching is done, occurs with
same probability as the
corresponding initial state.
But the
probability of the initial state
is the  probability the Born rule assigns to the full set of 
branch results which the final state acquired over history.
It follows that for any experiment which
produces macroscopic results and thus, according to our hypothesis
permanent records, the present 
theory will assign the same probability to each
possible result as does the Born rule.

Some randomly chosen $|\psi_i(t)>$ becomes what the
real world is made out of.
We obtain a kind of hidden variable theory,
with the hidden variables present in the initial $|\psi_{in, i}>$.
They emerge in macroscopic reality only over time through their
influence on the sequence of branching results
unfolding as $t$ progresses.

\section{\label{sec:stern}Stern-Gerlach Experiment}

We now apply the theory of Section \ref{sec:dqm} to a simplified model
of a Stern-Gerlach device. In addition to a particle interacting
with a magnetic field, we will need some degree
of freedom to serve as the environment entangled with the experimental results 
and thereby causing branching. 
A second particle will serve that purpose.

\subsection{\label{subsec:2dhilbertspace} Two Dimensional Lattice Fermions}

We assume a 2-dimensional
lattice in place of the 1-dimensional lattice of Section \ref{sec:complexity}.
The state space for the system $\mathcal{H}$ becomes
a tensor product of fermion spaces $\mathcal{H}_z$ one for each lattice point $(z_x, z_y)$
\begin{equation}\label{tensorproduct1}
\mathcal{H} = \otimes_z \mathcal{H}_z.
\end{equation}
As in Section \ref{sec:complexity}, an orthogonal basis for each $\mathcal{H}_z$ is the vacuum
$| 0>_z$ and 1-particle states $|1>_z, |-1>_z$ and the 2-particle state $|2>_z$.
The overall system vacuum $| 0>$ is the product
$\otimes_z |0>_z$. 

To adapt the definition of complexity in Section \ref{sec:complexity}, in Eq. (\ref{defk}) defining
$K$, in addition to hermitian
operators on $H_{z_0} \otimes H_{z_1}$ for $z_0$ and $z_1$ with $z_{0 x}$ and $z_{1 x}$
nearest neighbors,
we include also operators for which $z_{0 y}$ and $z_{1 y}$  are nearest neighbors. 
The remainder 
of Section \ref{sec:complexity} and Section \ref{sec:branching} can be adapted similarly.
The linear lower bounds of Appendices \ref{app:entangledpoint} and \ref{app:entangledextended}
can be adapted by interpreting
the present 2-dimensional lattice of spaces $\mathcal{H}_{(z_x, z_y)}$ as a 1-dimensional
lattice of spaces
\begin{equation}
\label{adapt}
\mathcal{H}_{z_y} = \otimes_{z_x} \mathcal{H}_{(z_x, z_y)}.
\end{equation}

It is convenient
to introduce a value of lattice spacing $a$ small with respect
to other length scales we will consider.
Except for instants when the Stern-Gerlach device
turns on a magnetic field, we assume
a non-interacting Hamiltonian which, acting on any state,
gives the sum of the non-relativistic kinetic energies
of any particles in that state.

For momentum vector $k = (k_x, k_y)$, let
$\psi(k, z_{in}, z, t)$
be a Gaussian wave function
with $z_x$ and $z_y$ coordinate means 
\begin{subequations}
\begin{eqnarray}
\label{psiin1}
\bar{z}_x(t) & = & \int d z_x \int d z_y z_x |\psi(k, z_{in}, z, t)|^2, \\
\label{psiin2}
\bar{z}_y(t) & = & \int d z_x \int d z_y z_y |\psi(k, z_{in}, z, t)|^2, \\
\label{psiin3}
\bar{z}_x(t) & = & \frac{k_x t}{m} + z_{in x} , \\
\label{psiin4}
\bar{z}_y(t) & = & \frac{k_y t}{m} + z_{in y}
\end{eqnarray}
\end{subequations}
and coordinate dispersions
\begin{subequations}
\label{disps1}
\begin{multline}
d_{z_x}(t)^2  =  \\ 
\int d z_x \int d z_y [z_x - \bar{z}_x(t)]^2 |\psi(k, z_{in}, z, t)|^2 ,
\end{multline}
\label{disps2}
\begin{multline}
d_{z_y}(t)^2  =  \\
\int d z_x \int d z_y [z_y - \bar{z}_y(t)]^2 |\psi(k, z_{in}, z, t)|^2 ,
\end{multline}
\label{disps3}
\begin{multline}
d_{z_x}(t)^2 = \frac{t^2}{4 d^2 m^2} + d^2 ,\hphantom{dddddddddddddddddddd} 
\end{multline}
\begin{multline}
\label{disps4}
d_{z_y}(t)^2 = \frac{t^2}{4 d^2 m^2} + d^2. \hphantom{dddddddddddddddddddd}
\end{multline}
\end{subequations}
Here $m$ is the particle mass, $d$ is chosen to set the initial
state dispersion, and we are using continuum expressions for the
means and dispersions in place of their more complicated discrete versions.

Define the 2-particle state $| z_0, s_0, z_1, s_1>$ to be
\begin{equation}\label{def2part}
| z_0, s_0, z_1, s_1> = |s_0>_{z_0} \otimes |s_1>_{z_1} \otimes_{z \ne z_0, z_1} | 0>_z. 
\end{equation}
and let $|k_0, z_{in 0}, s_0, k_1, z_{in 1}, t>$ be
\begin{multline}\label{def2part1}
|k_0, z_{in 0}, s_0, k_1, z_{in 1}, s_1, t> = \\
\sum_{ z_0 z_1} a^4 \psi( k_0, z_{in 0}, z_0, t) \psi(k_1, z_{in 1}, z_1, t) \times \\
| z_0, s_0, z_1, s_1>
\end{multline}

\subsection{\label{subsec:initialstate} Initial State}

With these various definitions in place, assume a 2-particle initial state at $t_{in}$
of 
\begin{multline}\label{psit0}
|\psi, t_{in}> =  
\frac{1}{\sqrt{2}}|k_0, z_{in 0}, 1, k_1, z_{in 1}, -1, t_{in}> - \\
\frac{1}{\sqrt{2}}|k_0, z_{in 0}, -1, k_1, z_{in 1}, 1, t_{in}>, 
\end{multline}
with momenta and initial positions
\begin{subequations}
\begin{eqnarray}\label{kandzinit1}
k_0 & = & (q, 0), \\
k_1 & = & (-q, 0), \\
z_{in 0} & = & (w,0), \\
z_{in1} & = & (-w,0).
\end{eqnarray}
\end{subequations}

This state is entangled at birth. Depending on the size of $b$
in relation to the position $w$ and the dispersion $d$, it follows 
from Sections \ref{sec:complexity} and \ref{sec:branching} that the optimal 
branch configuration may already split the state. Since the spin state is
rotationally invariant, however, there are actually a family of branch configurations
all of which would yield the same value for $Q(\{|\psi_i>\})$. By breaking
rotational symmetry by some arbitrarily small amount, a particular direction of branching
can be favored. However, whatever branching occurs at $t_{in}$ will
recombine and be replaced by a new pattern when the magnetic field
is applied, the result of which will be a single branch configuration
with far lower $Q(\{|\psi_i>\})$ than any other choice.

\subsection{\label{subsec:scattering} Scattering Event}

Now suppose at some later time $t_1$ a term is added to the Hamiltonian
for a very short time interval
causing states of particle 0 with spin 1 suddenly to
acquire a momentum component $k_{0 y}$ of 
$r$ and causing states with spin -1 suddenly to acquire a 
momentum component $k_{0 y}$ of $-r$. For $t > t_1$ the state becomes
\begin{multline}\label{psit1}
|\psi, t> = 
\frac{1}{\sqrt{2}}|k_{0 +}, z_{in 0 +}, 1, k_1, z_{in 1}, -1, t> - \\
\frac{1}{\sqrt{2}}|k_{0 -}, z_{in 0 -}, -1, k_1, z_{in 1}, 1, t> 
\end{multline}
with the new values
\begin{subequations}
\begin{eqnarray}\label{kandz}
k_{0 +} & = & (q, r), \\
k_{0 -} & = & (q, -r), \\
z_{in 0 +} & = & (w, -\frac{r t_1}{m}), \\
z_{in 0 -} & = & (w, \frac{r t_1}{m}).
\end{eqnarray}
\end{subequations}

This state has both entanglement between particles 0 and 1, and entanglement
between the spin 1 and spin -1 states of particle 1.
But it follows from Eqs. (\ref{psiin1}-\ref{psiin4}) that the spin 1 and
spin -1 components of particle 0 will separate in the $y$ 
direction with distance growing as $\frac{2 r t}{m}$. The dispersion in this separation
will grown as $\frac{t}{d m \sqrt{2}}$. If $r > \frac{1}{2 \sqrt{2} d}$ the two spin
components of particle 0 will completely separate as $t$ grows, and will therefore, if
left entangled in a single branch, contribute complexity larger than any finite $b$. The 
condition on $r$ for this to happen is simply that the $y$ momentum delivered at time
$t_1$ be larger than the wavefunction's $y$ momentum dispersion. This should certainly 
be fulfilled for any macroscopic Stern-Gerlach device.

\subsection{\label{subsec:finalinitial} Final State Ensemble, Initial State Ensemble}

The end result is that at some large $t_{out}$, the ensemble of branches which minimize $Q$ 
will asymptotically approach
\begin{subequations}
\begin{multline}
\label{psibranches}
|\psi, t_{out}>_0 = \\
|k_{0 +}, z_{in 0 +}, 1, k_1, z_{in 1}, -1, t_{out}>, 
\end{multline}
\begin{multline}
\label{psibranches1}
|\psi, t_{out}>_1  = \\
-|k_{0 -}, z_{in 0 -}, -1, k_1, z_{in 1}, 1, t_{out}>.
\end{multline}
\end{subequations}
Each branch will have weight $\frac{1}{2}$. In other words
the Stern-Gerlach device will yield spin 1 and -1
each with probability $\frac{1}{2}$. Which is the expected result
for this particular Stern-Gerlach model.

Reversing time
evolution on this ensemble back to $t_{in}$ to form an initial state
ensemble, then taking the limit of the branch decomposition time $t_{out} \rightarrow \infty$
gives
\begin{subequations}
\begin{eqnarray}
\label{psibranches2}
|\psi, t_{in}>_0 & = &
|k_0, z_{in 0}, 1, k_1, z_{in 1}, -1, t_{in}>, \\ 
\label{psibranches3}
|\psi, t_{in}>_1 & = &
-|k_0, z_{in 0}, -1, k_1, z_{in 1}, 1, t_{in}>.
\end{eqnarray}
\end{subequations}
Each of which again has weight $\frac{1}{2}$. The center
of the wave packet of the first of these states, at $t_1$ will
move purely in the positive $y$ direction. The center of the
wave packet for the second will move purely in the negative
$y$ direction. 

Thus each of the initial states, from the perspective of an
approximate macroscopic description, will show classical behavior.
The initial state vectors, in effect, carry hidden degrees
of freedom determining the result of the Stern-Gerlach experiment,
but invisible to the macroscopic world until a magnetic field is applied
at time $t_1$. 

Unlike the account of the experiment provided by environmentally
induced decoherence by itself, however, the account provided by unitary time evolution
of the initial state vectors applies through the entire process.
There is no missing beat before something real emerges.
The macroscopic description of the process remains approximate.
But underlying the approximate macro description, the microscopic
substance of reality is in place the whole time.

The split into branches of the state $|\psi, t_{out}>$ at some large $t_{out} > t_1$ in Eq. (\ref{psit1}),
done without the limit $t_{out} \rightarrow \infty$, would yield a $t_{in}$ ensemble differing from the
limiting states in Eqs. (\ref{psibranches1}, \ref{psibranches2}) by some tiny amount depending on $t_{out}$. 
Taking the splitting time $t_{out} \rightarrow \infty$, however, was already 
imposed by the goal of obtaining a branch ensemble with simple
properties under Lorentz transformation.

Although in the present simplified Stern-Gerlach model complexity
sufficient to cause branching of the final state arises 
from entanglement of
the scattered wave packets across an increasing distance, 
this is probably not how branching would occur in a more realistic model.
A more realistic model would include 
a large set of additional records of the scattering results
copied into the environment beyond the scattered particles
\cite{ Zeh, Zurek, Riedel}.
The complexity arising from this set of records would then
drive branching.

\section{\label{sec:bell}Bell's Theorem}

We now consider the application of Bell's theorem to 
the deterministic formulation of quantum mechanics in Section \ref{sec:dqm}.
Bell's theorem asserts, essentially, that a particular class
of local hidden variable theories, applied
to the measurement of the correlation of a pair
of spin-$\frac{1}{2}$ particles in a total angular momentum 0
state necessarily yields
predictions in contradiction to the predictions
of standard quantum mechanics.
But the theory of Section \ref{sec:dqm} is not local
in the sense required for Bell's theorem. 
Also, a key element of Bell's theorem is for the
spin measurements entering the correlation to be
performed with spacelike separation according
to spin direction choices also made at a spacelike separation.
The theory of Section \ref{sec:dqm}, however, allows no
freedom in setting up spin measurement devices.
Anything that occurs over history in this theory
must be programmed into the initial state vector at time $t_{in}$.
Despite these caveats, it may still
be useful to show that the theory of Section \ref{sec:dqm}
reproduces the correlations
expected of standard quantum mechanics in a Bell's theorem experiment.
Which we will now do.

\subsection{\label{Four Particle Initial State} Four Particle Initial State}

We return to the 1-dimensional lattice fermion theory of Section \ref{sec:complexity}.
We assume
4 particles in the initial state. Two to form the 
correlated angular momentum 0 state and two more to serve as
measuring devices to record the correlated spins.

As in Section \ref{sec:stern}, we
introduce a value of lattice spacing $a$ small with respect
to other length scales we will consider.
We assume
a Hamiltonian given by the
sum of kinetic energies of particles in that state,
along with a pair of infinite potential barriers details
soon to be specified.

For momentum $k$, let
$\psi(k, x_{in}, x, t)$
be a Gaussian wave function
with coordinate means and dispersions
given by the 1-dimensional version of Eqs. (\ref{psiin1}-\ref{disps4}).
Define the 4-particle state $| k_0, x_{in 0}, s_0,... k_3, x_{in 3}, s_3, t>$ to be
the 1-dimensional 4-particle analog of the state in Eqs.( \ref{def2part} - \ref{def2part1}).

We assume a 4-particle initial state at $t_{in}$
of
\begin{multline}\label{psit01d}
|\psi, t_{in}> = 
\sum_{s_0, ... s_3} \psi_{01}( s_0, s_1) \psi_2( s_2) \psi_3( s_3) \times \\
| k_0, x_{in 0}, s_0, ... k_3, x_{in 3}, s_3, t_{in}> 
\end{multline}
with spin wave functions
\begin{subequations}
\begin{eqnarray}
\label{spin1}
\psi_{01}( 1, 1) & = & 0, \\
\label{spin2}
\psi_{01}( 1, -1) & = & \frac{1}{\sqrt{2}}, \\
\label{spin3}
\psi_{01}( -1, 1) & = & -\frac{1}{\sqrt{2}}, \\
\label{spin4}
\psi_{01}( 1, 1) & = & 0, \\
\label{spin5}
\psi_2( 1) & = & 1, \\
\label{spin6}
\psi_2( -1) & = & 0, \\
\label{spin7}
\psi_3( 1) & = &  \cos(\frac{\theta}{2}), \\
\label{spin8}
\psi_3( -1) & = & \sin(\frac{\theta}{2})
\end{eqnarray}
\end{subequations}
and momenta and initial positions
\begin{subequations}
\begin{eqnarray}
\label{k0}
k_0 & = & q, \\
\label{k1}
k_1 & = & -q, \\
\label{k2}
k_2 & = & -q, \\
\label{k3}
k_3 & = & q, \\
\label{x0}
x_{in 0} & = & w, \\
\label{x1}
x_{in 1} & = & -w, \\
\label{x2}
x_{in 2} & = & 3w, \\
\label{x3}
x_{in 3} & = & -3w.
\end{eqnarray}
\end{subequations}
We assume the momentum $q$ is much larger than the momentum dispersion $\frac{1}{2 d}$
in Eq. (\ref{disps3}), and the initial positions $w$, $3 w$, are much larger than
the position dispersion $d$ in Eq. (\ref{disps3}).

As a consequence of the choice of initial positions and momenta, particles
0 and 2 will collide at $x$ of $2 w$ at time $\frac{m w}{q}$, where $m$ is the particle mass. At
the same time particles 1 and 3 will collide at $x$ of $-2 w$. 

\subsection{\label{subsec:twoevents} Two Scattering Events}

For the collision
between 0 and 2, we assume an infinite potential barrier is encountered for
the components of each with spin 1 and no interaction occurs
for other combinations of components.
The spin 1 component
of 0 is thus scattered back with momentum $-q$ and the corresponding component
of particle 2 is scattered forward with momentum $q$. Other combinations
of spin components pass through unaffected.
Thus the final momentum of particle 2 measures the spin 1
component of particle 0.

We assume a corresponding event for collision of particles 1 and 3 at position $x$
of $-2 w$ but with scattering for each particle in a state with spin wave function
$[\cos( \frac{ \theta}{2}), \sin( \frac{ \theta}{2})]$  and transparency
for the orthogonal state with spin wave function
$[-\sin( \frac{ \theta}{2}), \cos( \frac{ \theta}{2})]$. 
The final momentum of particle 3 thus
measures the component of particle 1 with spin wave function  
$[\cos( \frac{ \theta}{2}), \sin( \frac{ \theta}{2})]$.

We assume that the potential barriers are then turned off for $t$ beyond the collision
time $\frac{ m w}{q}$ by a margin of, say, $\frac{ m w}{ q}$ to
avoid additional unwanted scattering events.

As a result of the collision between particles 0 and 2 and the collision between 
particles 1 and 3, the states of each these pairs become entangled. And as
was the case for the entangled states of the Stern-Gerlach experiment of Section \ref{sec:stern},
the consequence of these entanglements is that the total state complexity will then rise linearly with $t$. 
For any finite value of b, for some large $t_{out}$, the minimum of $Q$ will occur for
a set of 4 branches $|e_0, e_1, t_{out}>$.
The index $e_0$ takes values of u(p) or d(own) specifying whether
the scattering of particles 0 and 2 gives a value of 1 or -1
for the spin of particle 0.  
The index $e_1$ takes values of u(p) or d(own) specifying whether
the scattering of particles 1 and 3 
gives a spin state for particle 1 with wave function 
$[\cos( \frac{ \theta}{2}), \sin( \frac{ \theta}{2})]$ 
or the orthogonal state with spin wave function 
$[-\sin( \frac{ \theta}{2}), \cos( \frac{ \theta}{2})]$.
The branch states are given by
\begin{multline}
\label{branchbell}
|e_0, e_1, t_{out}> = \eta_{e_0 e_1} \sum_{s_i} \psi_{e_0 e_1}( s_0,... s_3) \times \\
| k_{0 e_0}, x_{in e_0 0}, s_0, ... k_{3 e_1}, x_{in e_1 3}, s_3, t_{out}>, 
\end{multline}
where $| k_0, x_{in 0}, s_0,... k_3, x_{in 3}, s_3, t_{out}>$ is
the 1-dimensional 4-particle analog of the state in Eqs.( \ref{def2part} - \ref{def2part1}).
The spin wave function in Eq. (\ref{branchbell}) is
\begin{equation}
\label{spinbell}
 \psi_{e_0 e_1}( s_0,... s_3) =  
\psi_{e_0 0}( s_0) \psi_{e_1 1}( s_1) \psi_2( s_2) \psi_3( s_3), 
\end{equation}
where $\psi_2( s_2)$ and $\psi_3( s_3)$ are the same as given in 
Eqs. (\ref{spin5} - \ref{spin8}), 
$\psi_{u 0}( s_0)$ and $\psi_{u 1}( s_1)$ are the same as $\psi_2( s_0)$ and $\psi_3( s_1)$, 
respectively, and $\psi_{d 0}( s_0)$ and $\psi_{d 1}( s_1)$ are the vectors orthogonal to 
$\psi_{u 0}( s_0)$ and $\psi_{u 1}( s_1)$, respectively.
The momenta $k_{d 0}, ... k_{d 3}$ in Eq. (\ref{branchbell}) are the same as
$k_0, ... k_3$ in in Eqs. (\ref{k0} - \ref{k3})
while the momenta $k_{u 0}, ... k_{u 3}$ are $k_{d 0}, ... k_{d 3}$,
respectively, with signs flipped.
The initial positions $x_{in e_0 0}, ... x_{in e_1 3}$ have values
which, according Eq. (\ref{psiin3}),
put the mean positions $\bar{x}_0( t), ... \bar{x}_3(t)$ at the
corresponding collision points at $t$ of $\frac{m w}{ q}$.
Finally, the normalizations $\eta_{e_0 e_1}$ 
are such that the weights of these 4 branches become 
\begin{subequations}
\begin{eqnarray}
\label{upup}
< u, u, t | u, u, t> & = & \frac{1}{2} \sin^2( \frac{ \theta}{2}), \\ 
\label{updown}
< u, d,t | u, d, t> & = & \frac{1}{2} \cos^2( \frac{ \theta}{2}), \\
\label{downup}
< d, u,t | d, u, t> & = & \frac{1}{2} \cos^2( \frac{ \theta}{2}), \\
\label{downdown}
< d, d,t | d, d, t> & = & \frac{1}{2} \sin^2( \frac{ \theta}{2}).
\end{eqnarray}
\end{subequations}

\subsection{\label{subsec:result} Ensemble of Replicas}

Now replicate this prototype experiment $N$ times over displaced
sufficiently in space and time for distinct copies not to 
interfere.  For this combined system at large $t_{out}$, the
minimal value of $Q$ will occur for a set of $4^N$ branches
$| e_0^0, e_1^0, ...e_0^{N-1}, e_1^{N-1},t_{out}>$ 
with weights
\begin{multline}
\label{bigsys}
< e_0^0, e_1^0, ...e_0^{N-1}, e_1^{N-1},t_{out} | e_0^0, e_1^0, ...e_0^{N-1}, e_1^{N-1},t_{out}> \\
 = [\frac{1}{2}\sin^2(\frac{\theta}{2})]^{\sum_i a_i} [\frac{1}{2}\cos^2(\frac{\theta}{2})]^{\sum_i d_i},
\end{multline}
where $a(gree)_i$ is 1 for $i$ for which
$e_0^i$ and $e_1^i$ are $u$ and $u$ or $d$ and $d$, respectively, 
and 0 otherwise, and $d(isagree)_i$ is 
1 for $i$ for which
$e_0^i$ and $e_1^i$ are $u$ and $d$ or $d$ and $u$ respectively,
and 0 otherwise. 

For a branch randomly chosen from the ensemble in Eq. (\ref{bigsys}),
consider the average of $a_i - d_i$ over its copies
of the experiment. The result will be that branch's estimate for the
average of the product of observed values of the two spins. 
It follows from Eq. (\ref{bigsys}) that the result may be
viewed as the average of $N$ independent choices of a random
variable that takes the value 1 with probability $\sin^2(\frac{\theta}{2})$
and takes the value -1 with probability $\cos^2(\frac{\theta}{2})$.
By the central limit theorem, for large $N$ we expect any randomly
chosen branch to give nearly the result
\begin{subequations}
\begin{eqnarray}
\label{spindots}
\frac{1}{N} \sum_i (a_i - d_i) & = & \sin^2(\frac{\theta}{2}) - \cos^2(\frac{\theta}{2}) \\
& = & -\cos( \theta),
\end{eqnarray}
\end{subequations}
which is the expected result from standard quantum mechanics.
This is the result which Bell's theorem rules out for its class of local hidden variable theories.

\section{\label{sec:conclusion}Conclusion}

In Section \ref{sec:problems} we argued
that the branching which follows from environmentally induced decoherence
by itself looks like its missing something. The conjectures in 
Section \ref{subsec:branchingatinfinity}
propose to fill in what's missing. 

Multiple questions follow. 
What more can be said to support or refute
the hypothesis that macroscopic reality consists of
the accumulating set of persistent branching results?
What are the consequences, or possible ways to handle,
the limitation that $t_{out} \rightarrow \infty$ in Section \ref{sec:dqm}
ignores a cosmological time scale?
Does the lattice definition of net
complexity in Section \ref{subsec:branchcomplexity}, or something close to it, have a continuum limit,
and if it does, is the branch structure which it gives rise to
Lorentz covariant in the sense proposed in 
Section \ref{subsec:branchingatinfinity}? 
Would the continuum limit differ for 
Hermitian operators entering Eq. (\ref{defk}) allowed
to span more than 2 lattice sites? Are there other constants
which might show up in addition to b
if more complicated choices of operators are allowed in 
Eq. (\ref{defk})? 

At least for free lattice field theories
and for simple interacting theories,
it seems some of these questions might be answerable. There is
now a large amount of mathematically rigorous work
on the continuum limits of lattice field theories which
might be drawn on. Generally, what has emerged 
is that an infinite range of possible lattice theories
all reduce to some much smaller set of continuum theories 
often specified by only a finite dimensional space of parameters.

Finally, how could $b$ be measured? From multiple non-interacting 
copies of some single system, as in Section \ref{sec:bell}, 
the mean value of an observed complexity, via Eq. (\ref{defQ}),
might be used to find $b$. Among the problems with this proposal,
however, are that it needs to be done in the limit $t_{out} \rightarrow \infty$.

\begin{acknowledgments}
Thanks to Jess Riedel for an extended debate over an earlier
version of this work and to Cristi Stoica
for telling me about the work in \cite{Stoica}.
\end{acknowledgments}

\appendix

\section{\label{app:entangledpoint} Complexity of Entangled Particles at Distant Points}

The proof of Eq. (\ref{cbound1}) proceeds as follows. 
The trajectories $k(t) \in K$ and $U_k(t)$
which determine any $C( |\omega>, |\psi>)$, according to 
Eqs. (\ref{udot}) - (\ref{complexity}), we characterize 
by a corresponding
set of trajectories of Schmidt spectrum vectors. We then
find the rotation matrices which govern the motion of these vectors
as $t$ changes. A 
bound on the time integral of the angles which 
occur in these matrices by a time integral of $\parallel k(t) \parallel$ 
yields Eq. (\ref{cbound1}).

\subsection{\label{subsec:schmidtspectra} Schmidt Spectra}

For a trajectory $k(t) \in K$, let $U_k(t)$ be the solution to Eqs. (\ref{udot}) and (\ref{uboundary0}).
Define $|\psi(t)>$ to 
\begin{equation}
\label{phioft}
|\psi( t)> = U_k(t)|\psi>,
\end{equation} 
for a 2 particle product state $|\psi>$, 
and assume that $k(t)$ has been chosen
to give
\begin{equation}
\label{upsiphi}
\xi |\psi(1)> = |\omega>,
\end{equation}
for some complex $\xi$ with $|\xi| = 1$. 

For any $p$ in the range $0 \le p < n$, split the state space
$\mathcal{H}$ into the product
\begin{subequations}
\begin{eqnarray}
\label{defq}
\mathcal{Q}_p & = & \otimes_{q \le p} \mathcal{H}_q, \\
\label{defr}
\mathcal{R}_p & = & \otimes_{q > p} \mathcal{H}_q, \\
\label{split}
\mathcal{H} & = & \mathcal{Q}_p \otimes \mathcal{R}_p.
\end{eqnarray}
\end{subequations}

Since the number operators 
$N_{\mathcal{Q}_p}$ on $\mathcal{Q}_p$ and $N_{\mathcal{R}_p}$ on $\mathcal{R}_p$
commute, 
$|\psi(t)>$ can be decomposed into a sum of terms each of which is an eigenvector of both
operators. The $N_{\mathcal{Q}_p}$, $N_{\mathcal{R}_p}$ eigenvalue pairs which occur
must sum to 2 since $|\psi(t)>$ is an eigenvector of the total particle number
operator $N$ with eigenvalue 2. 
The sum of the $\mathcal{Q}_p \otimes \mathcal{R}_p$
Schmidt decompositions of $N_{\mathcal{Q}_p}$, $N_{\mathcal{R}_p}$ eigenvectors then gives a 
Schmidt decomposition of $|\psi(t)>$, which can be arranged in the form
\begin{subequations}
\begin{eqnarray}
\label{schmidt}
|\psi(t) & = & \sum_i s_{ip}(t) |\phi_{ip}(t)>|\chi_{ip}(t)>, \\
\label{nq0}
N_{\mathcal{Q}_p} |\phi_{0p}(t)> & = & 0, \\
\label{nr0}
N_{\mathcal{R}_p} |\chi_{0p}(t)> & = & 2 |\chi_{0p}(t)>, \\
\label{nq1}
N_{\mathcal{Q}_p} |\phi_{1p}(t)> & = & 2 |\phi_{1p}(t)>, \\
\label{nr1}
N_{\mathcal{R}_p} |\chi_{1p}(t)> & = & 0, \\
\label{nq2}
N_{\mathcal{Q}_p} |\phi_{ip}(t)> & = & |\phi_{ip}(t)>, i \ge 2, \\
\label{nr2}
N_{\mathcal{R}_p} |\chi_{ip}(t)> & = & |\chi_{ip}(t)>, i \ge 2,
\end{eqnarray}
\end{subequations}
with $|\phi_{ip}> \in \mathcal{Q}_p$, $|\chi_{ip}> \in \mathcal{R}_p$,
all Schmidt spectrum values $s_{ip}(t)$ real nonnegative, and the $s_{ip}(t)$
in decreasing order for $i \ge 2$.
The normalization of $|\psi(t)>$ implies
\begin{equation}
\label{normalization}
\sum_i s_{ip}(t)^2 = 1.
\end{equation}

As $t$ goes from $0$ to $1$, the Schmidt spectrum vector $[ s_{0p}(t), s_{1p}(t), ... ]$ follows
a smooth trajectory across a (high dimensional) unit sphere.
A lower bound can be placed on the total angular rotation of
$[ s_{0p}(t), s_{1p}(t), ... ]$. 

The initial Schmidt vector $[ s_{0p}( 0), s_{1p}(0), ...]$, since it
is from 
a 2 particle product state $|\psi( 0)> = |\psi>$, 
has the form $[ s_{0p}(0), s_{1p}(0), s_{2p}(0), 0, ...]$ with at most 3 nonzero entries.
The final Schmidt vector $[ s_{0p}( 1), s_{1p}(1), ...]$,
according to Eqs. (\ref{omega0}) - (\ref{omega}), has the
form $[ 0, 0, \frac{1}{\sqrt{2}}, \frac{1}{\sqrt{2}}, 0, ...]$ with exactly 2 nonzero entries.
Thus over the interval from $t$ of $0$ to $1$, $[ s_{0p}(t), s_{1p}(t), ... ]$ must
go through a sequence of rotations with a total angle of at least $\frac{\pi}{4}$.  For the small time interval
from $t$ to $t + \delta t$, let $u_{ip}(t)$ and $\theta_p(t)$ be 
\begin{subequations}
\begin{eqnarray}
\label{sdeltat}
s_{ip}( t + \delta t) & = & s_{ip}( t) + \delta t u_{ip}(t), \\
\label{uoft}
[ \theta_p( t)]^2 & = & \sum_i [ u_{ip}( t)]^2. 
\end{eqnarray}
\end{subequations}
We then have
\begin{equation}
\label{thetabound}
\sum_{0 \le p < n} \int_0^1 | \theta_p(t)| d t \ge \frac{n \pi}{4}.
\end{equation}

\subsection{\label{subsec:schmidtrotation} Schmidt Rotation Matrix}

The rotation of $[ s_{0p}(t), s_{1p}(t), ... ]$ during the interval from $t$ to $t + \delta t$
will be determined by $k(t)$. In Eq. (\ref{defk}), the only term which will give rise to 
a nonzero value for $\theta_p(t)$ is $\hat{g}_{p(p+1)}(t)$. All other terms in Eq. (\ref{defk})
will either contribute a unitary transformation on $\mathcal{Q}_p$ and identity on $\mathcal{R}_p$
or an identity on $\mathcal{Q}_p$ and unitary on $\mathcal{R}_p$. Either of these two
alternatives will leave $[ s_{0p}(t), s_{1p}(t), ... ]$ unchanged.
The effect of $\hat{g}_{p(p+1)}(t)$ on $[ s_{0p}(t), s_{1p}(t), ... ]$ over the
interval from $t$ to $t + \delta t$ can therefore be determined from the simplification
\begin{equation}
\label{psisimp}
|\psi(t + \delta t)> = \exp[ i \delta t \hat{g}_{p(p+1)}(t)] |\psi(t)>.
\end{equation}

From $|\psi(t + \delta t)>$ of Eq. (\ref{psisimp}),
construct the density operator $\rho_p(t + \delta t)$ by 
a partial trace over $\mathcal{R}_p$, 
for which it is convenient to use the basis for $\mathcal{R}_p$
taken from the Schmidt decomposition of $|\psi(t)>$ at the beginning of the time 
interval
\begin{multline}
\label{defrho}
\rho_p(t + \delta t) = \\
\sum_i < \chi_{ip}(t)|\psi(t + \delta t)><\psi(t + \delta t)|\chi_{ip}(t)>.
\end{multline} 
An eigenvector decomposition of $\rho_p(t + \delta t)$ exposes
the vector $[ s_{0p}(t + \delta t), s_{1p}(t + \delta t), ... ]$
\begin{multline}
\label{rhodeltat}
\rho_p(t + \delta t) = \\
\sum_i [s_{ip}(t + \delta t)]^2 |\phi_{ip}(t + \delta t)><\phi_{ip}(t + \delta t)|.
\end{multline}

A bound state perturbation expansion to leading order in 
small $\delta t$ applied to Eqs. (\ref{psisimp}), (\ref{defrho}) and (\ref{rhodeltat})
then gives for $u_{ip}(t)$ of Eq. (\ref{sdeltat})
\begin{equation}
\label{ufromperturb}
u_{ip}(t) = \sum_j r_{ijp}(t) s_{jp}(t), 
\end{equation}
for the rotation matrix $r_{ijp}(t)$
\begin{multline}
\label{rijp}
r_{ijp}(t) = 
 -\operatorname{Im}( <\phi_{ip}(t)|<\chi_{ip}(t)| \\
\hat{g}_{p(p+1)}(t)|\phi_{jp}(t)>|\chi_{jp}(t)>).
\end{multline}

\subsection{\label{subsec:anglebounds} Rotation Angle Bounds}

Since $\hat{g}_{p(p+1)}(t)$ preserves $N_p + N_{p+1}$,
it is useful to decompose it into 
a linear combination of an operator $\hat{z}^0_{p(p+1)}(t)$
which, acting on $N_p + N_{p+1}$ eigenvectors,
gives nonzero results only for eigenvalue 0, 
and an operator
$\hat{z}^1_{p(p+1)}(t)$ which, acting on
$N_p + N_{p+1}$ eigenvectors,
gives nonzero results only for $N_p + N_{p+1}$ eigenvalues other than 0
\begin{multline}
\label{splitg}
\hat{g}_{p(p+1)}(t) = \\ a^0_{p(p+1)}(t) \hat{z}^0_{p(p+1)}(t) +
a^1_{p(p+1)}(t) \hat{z}^1_{p(p+1)}(t).
\end{multline}
Both operators we assume normalized to 1
\begin{equation}
\label{normgm}
\parallel \hat{z}^i_{p(p+1)}(t) \parallel = 1.
\end{equation}
It follows that $\hat{z}^0_{p(p+1)}(t)$ is constant 
over $t$ and
given by the projection operator
\begin{multline}
\label{zprojection}
\hat{z}^0_{p(p+1)}(t) = \\ |0>_p|0>_{p+1}<0|_p<0|_{p+1} \otimes_{q \ne p, p+1} I_q.
\end{multline}

Combining Eqs. (\ref{uoft}), (\ref{ufromperturb}) - (\ref{splitg}) gives
\begin{subequations}
\begin{eqnarray}
\label{thetasum}
| \theta_p(t)| & \le & |\theta_p^0( t)| + |\theta_p^1( t)|, \\
\label{thetasupi}
|\theta_p^i(t)|^2 & = & \sum_j [u_{jp}^i(t)]^2, 
\end{eqnarray}
\end{subequations}
with the definition
\begin{multline}
\label{usupi}
u_{jp}^i(t) =  -a^i_{p(p+1)}(t) \sum_k \operatorname{Im}\{ <\phi_{jp}(t)|<\chi_{jp}(t)| \\
\hat{z}^i_{p(p+1)}(t)|\phi_{kp}(t)>|\chi_{kp}(t)> s_{kp}(t)\}.
\end{multline}

For $u_{jp}^0(t)$ we then have
\begin{multline}
\label{usup0}
u_{jp}^0(t) =  a^i_{p(p+1)}(t) \sum_k \operatorname{Im}\{ <\phi_{jp}(t)|<\chi_{jp}(t)| \\
[I - \hat{z}^0_{p(p+1)}(t)]|\phi_{kp}(t)>|\chi_{kp}(t)> s_{kp}(t)\},
\end{multline}
since the set of $|\phi_{ip}(t)>|\chi_{jp}(t)>$ is an orthonormal
basis for $\mathcal{H}$ and therefore
\begin{multline}
\label{imis0}
\operatorname{Im}\{ <\phi_{jp}(t)|<\chi_{jp}(t)| \\
I|\phi_{kp}(t)>|\chi_{kp}(t)>\} = 0.
\end{multline}

But in addition
\begin{equation}
\label{schmidt1}
|\psi( t)> = \sum_k |\phi_{kp}(t)>|\chi_{kp}(t)> s_{kp}(t).
\end{equation}
Eqs. (\ref{zprojection}), (\ref{thetasupi}) - (\ref{schmidt1}), then give
\begin{multline}
\label{theta0bound}
[\theta_p^0(t)]^2 \le \\ [a^0_{p(p+1)}(t)]^2 < \psi(t)|[I - \hat{z}^0_{p(p+1)}(t)]|\psi(t)>.
\end{multline}

For $u_{jp}^1(t)$, since $\hat{z}^1_{p(p+1)}(t)$ is nonzero only on the
subspace with $N_p + N_{p+1}$ nonzero, we have
\begin{multline}
\label{usup1}
u_{jp}^1(t) =  -a^i_{p(p+1)}(t) \sum_k \operatorname{Im}\{ <\phi_{jp}(t)|<\chi_{jp}(t)| \\
\hat{z}^1_{p(p+1)}(t) [I - \hat{z}^0_{p(p+1)}(t)]|\psi(t)>\}.
\end{multline}
Eqs. (\ref{thetasupi}) and (\ref{usup1}) give
\begin{multline}
\label{theta1bound}
[\theta_p^1(t)]^2 \le [a^1_{p(p+1)}(t)]^2 < \psi(t)| 
[I - \hat{z}^0_{p(p+1)}(t)] \\
[\hat{z}^1_{p(p+1)}(t)]^2[I - \hat{z}^0_{p(p+1)}(t)]|\psi(t)>.
\end{multline}
But by Eq. (\ref{normgm}), the underlying operator $[z^1_{p(p+1)}(t)]^2$ on 
$\mathcal{H}_p \otimes \mathcal{H}_{p+1}$ has trace 1 and therefore all eigenvalues
bounded by 1. Thus Eq. (\ref{theta1bound}) implies
\begin{multline}
\label{theta1bound1}
[\theta_p^1(t)]^2 \le \\ 
[a^1_{p(p+1)}(t)]^2 < \psi(t)| [I - \hat{z}^0_{p(p+1)}(t)]|\psi(t)>.
\end{multline}

Eqs. (\ref{thetasum}), (\ref{theta0bound}) and (\ref{theta1bound1}) give
\begin{multline}
\label{thetafinal0}
\sum_p |\theta_p(t)| \le 
\sum_p \{ [|a^0_{p(p+1)}(t)| + |a^1_{p(p+1)}(t)|] \times
\\ \sqrt{ <\psi(t)| [I - \hat{z}^0_{p(p+1)}(t)]|\psi(t)>} \}.
\end{multline}
The Cauchy-Schwartz inequality then implies
\begin{multline}
\label{thetafinal}
[\sum_p |\theta_p(t)|] ^ 2 \le
\sum_p [|a^0_{p(p+1)}(t)| + |a^1_{p(p+1)}(t)|]^2 \times
\\ \sum_p <\psi(t)| [I - \hat{z}^0_{p(p+1)}(t)]|\psi(t)>.
\end{multline}

The state $|\psi(t)>$ can be expanded as a linear combination of orthogonal states 
each with 
2 particles, the first particle at some single point $x_0$, the second at a point $x_1$
which could coincide with $x_0$ if the spins are opposite. Each such state will survive the projection
$I - \hat{z}^0_{p(p+1)}(t)$ for at most 4 values of $p$, namely $x_0, x_0 - 1, x_1, x_1 - 1$. Thus
\begin{equation}
\label{psiprojectionbound}
\sum_p <\psi(t)| [I - \hat{z}^0_{p(p+1)}(t)]|\psi(t)> \le 4.
\end{equation}
In addition, by Eqs. (\ref{defk}), (\ref{defnormk}), (\ref{splitg}) and (\ref{normgm}),
\begin{multline}
\label{normkt}
\sum_p [|a^0_{p(p+1)}(t)| + |a^1_{p(p+1)}(t)|]^2 \le \\
2 \sum_p [|a^0_{p(p+1)}(t)|^2 + |a^1_{p(p+1)}(t)|^2]  \le \\
2 \parallel k(t) \parallel ^2.
\end{multline}

Assembling Eqs. (\ref{thetafinal}), (\ref{psiprojectionbound}) and (\ref{normkt}) gives
\begin{equation}
\label{summedtheta}
\sum_p |\theta_p(t)| \le 2 \sqrt{2} \parallel k(t) \parallel.
\end{equation}
Eqs. (\ref{complexity}), (\ref{thetabound}) and (\ref{summedtheta}) 
then yield 
\begin{equation}
\label{app:cbound}
C( |\omega>, |\psi>) \ge \frac{ n \pi}{ 8 \sqrt{2}}.
\end{equation}
and therefore 
\begin{equation}
\label{app:cbound1}
C( |\omega>) \ge \frac{ n \pi}{ 8 \sqrt{2}}.
\end{equation}

\subsection{\label{subsec:upperbound} Complexity Upper Bound}

An upper bound on $C( |\omega>)$ is given by $C( |\omega>, |\psi>)$ for any product state
$|\psi>$, which in turn is given by $\int_0^1 d t \parallel k( t) \parallel$ for any 
trajectory $k(t)$ connecting $|\psi>$ to $|\omega>$, one of which we now construct for
\begin{equation}
\label{psiupperbound}
|\psi> = |1>_0 |1>_1 \otimes_{q \ne 0, 1} |0>_q.
\end{equation}

Define $k_0$ to be
\begin{equation}
\label{defk0}
k_0 = -i (|a_0>< b_0| - |b_0><a_0|) \otimes_{q \ne 0, 1} I_q,
\end{equation}
where
\begin{subequations}
\begin{eqnarray}
\label{defpsi0}
|a_0>  & = & |-1>_0 |-1>_1, \\ 
\label{defpsi1}
|b_0>  & = & |1>_0 |1>_1.
\end{eqnarray}
\end{subequations}
In addition, for $1 \le i < n$, define $k_i$ to be
\begin{multline}
\label{defki}
k_i = -i (|a_i>< b_i| - |b_i><a_i| + \\ 
|c_i>< d_i| - |d_i><c_i| ) \otimes_{q \ne i, i+1} I_q,
\end{multline}
where
\begin{subequations}
\begin{eqnarray}
\label{defai}
|a_i> & = & |0>_i |1>_{i+1}, \\
\label{defbi}
|b_i> & = & |1>_i |0>_{i+1}, \\
\label{defci}
|c_i> & = & |0>_i |-1>_{i+1}, \\
\label{defdi}
|d_i>  & = &  |-1>_i |0>_{i+1}.
\end{eqnarray}
\end{subequations}

From these states, define the projection operator
\begin{equation}
\label{proj0}
P_0 = (|a_0><a_0| + |b_0><b_0|) \otimes_{q \ne 0, 1} I_q,
\end{equation}
and for $1 \le i < n$ the projections
\begin{multline}
\label{proji}
P_i  =  (|a_i><a_i| + |b_i><b_i| + \\
|c_i><c_i| +|d_i><d_i|) \otimes_{q \ne i, i + 1} I_q.
\end{multline}
We then have for any $i \ge 0$
\begin{equation}
\label{expki}
\exp( i \theta k_i) = (I - P_i) + \cos( \theta) P_i + i \sin( \theta) k_i.
\end{equation}
Define iteratively
\begin{subequations}
\begin{eqnarray}
\label{firststep}
|\psi_0> & = & \exp( i\frac{\pi}{4} k_0) |\psi>, \\
\label{nextstep}
|\psi_i> & = & \exp( i\frac{\pi}{2} k_i) |\psi_{i - 1}>, i \ge 1.
\end{eqnarray}
\end{subequations}
Eqs. (\ref{proj0}) - (\ref{expki}) then imply
\begin{equation}
\label{expk0kn}
|\psi_{n-1}> = |\omega>. 
\end{equation}

The norms of these operators are
\begin{subequations}
\begin{eqnarray}
\label{normk0}
\parallel k_0 \parallel & = & \sqrt{2}, \\
\label{normki}
\parallel k_i \parallel & = & 2, i \ge 1.
\end{eqnarray}
\end{subequations}
We therefore have
\begin{equation}
\label{cupper}
C( |\omega>, |\psi>) \le ( n - 1 + \frac{\pi}{2 \sqrt{2}}) \pi.
\end{equation}
Eq. (\ref{cupper}) then implies 
\begin{equation}
\label{app:cbound2}
C( |\omega>) \le (n - 1 + \frac{1}{2\sqrt{2}}) \pi.
\end{equation}

\section{\label{app:entangledextended} Entangled Extended State}

\subsection{\label{subsec:extendedlowerbound} Lower Bound on the Complexity of an Entangled Extended State}

Repeating Section \ref{subsec:schmidtspectra}, assume $k(t)$ and $U_k(t)$ 
solve Eqs. (\ref{udot}) and (\ref{uboundary0}) and connect some 2 particle product state $|\psi>$ to
$|\omega'>$ of Eqs. (\ref{omegaprime0}) - (\ref{omegaprime}) according to Eqs. (\ref{phioft}) and 
(\ref{upsiphi}). 

For any $p$ in the range $0 \le p < n + r$, split $\mathcal{H}$ into $\mathcal{Q}_p$ and 
$\mathcal{R}_p$ according to Eqs. (\ref{defq}) - (\ref{split}). The Schmidt decomposition of
$|\psi(t)>$ will again be given by Eqs. (\ref{schmidt}) - (\ref{nr2}), and
the initial Schmidt vector $[ s_{0p}( 0), s_{1p}(0), ...]$,  since it
is from 
a 2 particle product state $|\psi( 0)> = |\psi>$,
has, as before, the form $[ s_{0p}(0), s_{1p}(0), s_{2p}(0), 0, ...]$ with at most 3 nonzero entries. 
For $0 \le p < n$, the final Schmidt vector 
$[ s_{0p}( 1), s_{1p}(1), ...]$ remains $[ 0, 0, \frac{1}{\sqrt{2}}, \frac{1}{\sqrt{2}}, 0, ...]$ 
with exactly 2 nonzero entries, requiring again a total rotation angle of at least $\frac{\pi}{4}$. For
$\theta_p(t)$ of Eqs. (\ref{sdeltat}) and (\ref{uoft}), Eq. (\ref{thetabound}) is then repeated.

For $n \le p < n + r$, the initial Schmidt vector $[ s_{0p}( 0), s_{1p}(0), ...]$ still has at most
3 nonzero entries $[ s_{0p}(0), s_{1p}(0), s_{2p}(0), 0, ...]$.
But as a result of the extended wave functions in Eqs. (\ref{omegaprime0}) and 
(\ref{omegaprime1}), the final Schmidt vector $[ s_{0p}( 1), s_{1p}(1), ...]$
has 3 nonzero entries 
\begin{subequations}
\begin{eqnarray}
\label{s1p}
s_{1p}(1) & = & \sqrt{\frac{p - n + 1}{r}}, \\
\label{s2p}
s_{2p}(1) & = & \sqrt{\frac{r + n - p - 1}{2r}}, \\
\label{s3p}
s_{3p}(1) & = & \sqrt{\frac{r + n - p - 1}{2r}}.
\end{eqnarray}
\end{subequations}

The minimum total angle between initial and final Schmidt vectors becomes
$\arcsin( \sqrt{\frac{r + n - p - 1}{2r}})$, thus ranges from slightly less than $\frac{\pi}{4}$
for $p = n$ to slightly more than 0 for $p = n + r - 2$. 
Eq. (\ref{thetabound}) becomes
\begin{equation}
\label{thetaboundextended}
\sum_{0 \le p < n + r - 1} \int_0^1 | \theta_p(t)| d t \ge \frac{n \pi}{4} + r \kappa,
\end{equation}
where $\kappa$ is
\begin{equation}
\label{app:defkappa}
\kappa = \frac{1}{r} \sum_{0 < s < r} \arcsin( \sqrt{\frac{s}{2r}}).
\end{equation}
For large $r$, $\kappa$ approaches
\begin{equation}
\label{app:defkappalim}
\lim_{r \rightarrow \infty} \kappa = \int_{0}^{1} \arcsin( \sqrt{\frac{x}{2}}) d x.
\end{equation}
Eqs. (\ref{thetaboundextended}) and (\ref{summedtheta}) then give
\begin{equation}
\label{app:cbound1extended}
C( |\omega'>) \ge \frac{n \pi}{8 \sqrt{2}} + \frac{r \kappa}{2 \sqrt{2}}.
\end{equation}
The complexity of an entangled pair with one of the particles in an
extended state grows at least linearly both with the distance between
the pairs and with the width of the extended state.

\subsection{\label{subsec:extendedupperbound} Upper Bound on the Complexity of an Entangled Extended State}

An upper bound on $C(|\omega'>)$ is given by $C(|\omega'>, |\psi>)$ found by extending the trajectories
$k(t)$ and $U_k(t)$ of Section \ref{subsec:upperbound}. 

For $n \le i < n + r - 1$, starting from $|\psi_{n-1}>$
of Eq. (\ref{nextstep}), define 
\begin{subequations}
\begin{eqnarray}
\label{nextnextstep}
|\psi_i> & = & \exp( i \theta_i k_i) |\psi_{i - 1}>, \\
\label{sinthetai}
\sin( \theta_i) & = & \sqrt{ \frac{ n + r - i - 1}{n + r - i}}. 
\end{eqnarray}
\end{subequations}

Eqs. (\ref{proji}) and (\ref{expki}) imply
\begin{multline}
\label{psii}
|\psi_i>  = \sqrt{\frac{i + 1 - n}{2 r }} | \psi_{0i}> + \sqrt{\frac{i + 1 - n}{2r}} | \psi_{1i}> +\\
\sqrt{\frac{n + r -i - 1}{2r}} | \psi_{2i}> + \sqrt{\frac{n +r -i -1}{2r}} | \psi_{3i}>
\end{multline}
where
\begin{subequations}
\begin{multline}
\label{psi0i}
|\psi_{0i}>  = | -1>_0 \\ \otimes ( \frac{1}{\sqrt{i + 1 - n}} \sum_{ n \le x \le i }|-1>_x \otimes_{ q \ne 0, x} | 0>_q),
\end{multline}
\begin{multline}
\label{psi1i}
|\psi_{1i}> = | 1>_0 \\ \otimes ( \frac{1}{\sqrt{i + 1 - n}} \sum_{ n \le x \le i}|1>_x \otimes_{ q \ne 0, x} | 0>_q),
\end{multline}
\begin{multline}
\label{psi2i}
|\psi_{2i}>  =  | -1>_0|-1>_{i+1} \otimes_{ q \ne 0, i+1} | 0>_q,
\end{multline}
\begin{multline}
\label{psi3i}
|\psi_{3i}> = | 1>_0|1>_{i+1} \otimes_{ q \ne 0, i+1} | 0>_q. 
\end{multline}
\end{subequations}

We then have
\begin{equation}
\label{finalpsi}
|\psi_{n + r -2}> = |\omega'>.
\end{equation}
Combining Eqs. (\ref{normk0}), (\ref{normki}), (\ref{firststep}), (\ref{nextstep}), (\ref{nextnextstep}) and (\ref{sinthetai})
gives
\begin{equation}
\label{app:comegaprime}
C( |\omega'>, |\psi>) \le ( n - 1 + \frac{\pi}{2 \sqrt{2}}) \pi + 2 \lambda r,
\end{equation}
where
\begin{equation}
\label{app:defkappaprime}
\lambda = \frac{1}{r} \sum_{0 < s < r} \arcsin(\sqrt{ \frac{s}{s +1}}).
\end{equation}
For large $r$ for most of the range $0 < s < r$, the argument of the sum in Eq. (\ref{app:defkappaprime}) is nearly $\frac{\pi}{2}$.
Therefore
\begin{equation}
\label{app:limkappaprime}
\lim_{r \rightarrow \infty} \lambda = \frac{\pi}{2}.
\end{equation}
Eq. (\ref{app:comegaprime}) implies
\begin{equation}
\label{app:cupperprime}
C( |\omega'>) \le ( n - 1 + \frac{\pi}{2 \sqrt{2}}) \pi + 2 \lambda r.
\end{equation}

\section{\label{app:complexitygroup} Complexity Group}

We now show
that the group of $G$ all $U_k( 1)$ realizable as solutions to Eqs. (\ref{udot}) and
(\ref{uboundary0}) has as a subgroup the direct product
\begin{equation}
\label{formofga}
\hat{G} = \times_n SU(d_n),
\end{equation}
where $SU(d_n)$ acts on the subspace of $\mathcal{H}$
with eigenvalue $n$ of the total number operator $N$, 
$d_n$ is the dimension of this subspace, and
the product
is over the range $ 1 \le n \le 4 x_{max} + 1$.
Missing from this range of $n$ are only two states,
the vacuum, $n = 0$, and the state with every site occupied by
two particles, $n = 4 x_{max} + 2$.

For any integer $-x_{max} + 1 \le p \le x_{max}$,
divide $\mathcal{H}$ into the product
\begin{subequations}
\begin{eqnarray}
\label{defqa}
\mathcal{Q}_p & = & \otimes_{z \le p} \mathcal{H}_z, \\
\label{defra}
\mathcal{R}_p & = & \otimes_{z > p} \mathcal{H}_z, \\
\label{splita}
\mathcal{H} & = & \mathcal{Q}_p \otimes \mathcal{R}_p.
\end{eqnarray}
\end{subequations}

Let $K_p$ be the vector space of operators
given by Eq. (\ref{defk}) constructed from the set $\mathcal{F}_x$
of traceless particle number preserving Hermitian operators on
$\mathcal{H}_x$ with $x \le p$ and
the set $\mathcal{G}_{xy}$
of traceless particle number preserving Hermitian operators 
on nearest neighbor 
$\mathcal{H}_x \otimes \mathcal{H}_y$ orthogonal
to $\mathcal{F}_x$ and $\mathcal{F}_y$ with 
$x, y \le p$. Let $G_p$ be the group on 
$\mathcal{H}$
of all $U_k(1)$ realizable as solutions to Eq. (\ref{udot}) 
for $k(t) \in K_p$. 

The group
$G_p$ consists of all operators of the form
$\exp( i h)$ for $h \in L_p$, where $L_p$ is the 
Lie algebra generated by $K_p$ \cite{Divincenzo}.
Said differently,  $L_p$ is the smallest vector space of operators 
such that  $K_p \subseteq L_p$ and, in addition, for
any $h_0, h_1 \in L_p$, and any real $r_0, r_1$,
there are $h_2, h_3 \in L_p$ given by
\begin{subequations}
\begin{eqnarray}
\label{linear}
h_2 & = & r_0 h_0 + r_1 h_1, \\
\label{commute}
h_3 & = & i [ h_0, h_1].
\end{eqnarray}
\end{subequations}
The requirement that $L_p$ be closed under
sums in Eq. (\ref{linear}) follows from the Trotter product
formula applied to the large $t$ limit of
\begin{equation}
\label{linear1}
[ \exp( i t^{-1}r_0 h_0) \exp( i t^{-1} r_1 h_1)]^t.
\end{equation}
The requirement that $L_p$ be closed under commutation in
Eq. (\ref{commute}) follows from the 
Baker-Campbell-Hausdorff 
formula applied to the large $t$ limit of
\begin{multline}
\label{commute1}
[ \exp( i t^{-1/2} h_0) \exp( i t^{-1/2} h_1) \\ 
\exp( -i t^{-1/2} h_0) \exp( -i t^{-1/2} h_1)]^t.
\end{multline}

By induction on $p$, we will show that $G_p$
includes the subgroup $\hat{G}_p$
\begin{subequations}
\begin{eqnarray}
\label{formofgp}
\hat{G}_p & = & \times_n \hat{G}_{p n}, \\
\label{formofgpn}
\hat{G}_{p n} & = & SU(d_{p n}) \otimes_{z > p} I_z,
\end{eqnarray}
\end{subequations}
where $SU(d_{p n})$ acts on the subspace $\mathcal{Q}_{p n}$ of $\mathcal{Q}_p$
with eigenvalue $n$ of the total number operator $N$, and 
$d_{p n}$ is the dimension of $\mathcal{Q}_{p n}$. 
The product in Eq. (\ref{formofgp})
is over $1 \le n \le 2 x_{max} + 2 p + 1$.
Eqs. (\ref{formofgp})
(\ref{formofgpn}) for the case $p = x_{max}$ become Eq. (\ref{formofga}). 

For $p = -x_{max} + 1$, Eqs. (\ref{formofgp}) and Eqs. (\ref{formofgpn})
follow immediately from the definition of $K_p$. 
Assuming Eqs. (\ref{formofgp}) and Eqs. (\ref{formofgpn}) for
some $p - 1$, we will prove them for $p$.

Let $S_{p n}$ be an orthonormal basis for $\mathcal{Q}_{p n}$ consisting of
all vectors of the form
\begin{subequations}
\begin{eqnarray}
\label{psi0}
| \psi > & = & \otimes_{x \le p} |\psi(x)>_x, \\
\label{psi1}
\psi( x) & \in & \{ -1, 0, 1, 2 \}, \\
\label{psi2}
\sum_{x \le p} |\psi(x)| & = & n.
\end{eqnarray}
\end{subequations}
For any pair of distinct $|\psi_0>, |\psi_1> \in S_{p n}$, and
2 x 2 traceless Hermitian $h$, define
\begin{multline}
\label{defcaph}
H( |\psi_0>, |\psi_1>, h) = \\
\sum_{ij} |\psi_i><\psi_j| h_{ij} \otimes_{z > p} I_z.
\end{multline}
The set of all such $H( |\psi_0>, |\psi_1>, h)$ is a linear basis
for the Lie algebra $L_{p n}$ of the group $\hat{G}_{p n}$ of Eq. (\ref{formofgpn}). 

Thus to prove Eqs. (\ref{formofgpn}) and (\ref{formofgp}) for $p$ it is 
sufficient to show that any $H( |\psi_0>, |\psi_1>, h)$ for some $|\psi_0>, |\psi_1> \in S_{p n}$ and
2 x 2 traceless Hermitian $h$, given the induction hypothesis, is contained in the Lie algebra generated
by $L_{p-1 m}$ for some $m$ and 
all 
\begin{equation}
\label{pminus1p}
\hat{k} = k \otimes_{z \ne p-1, p} I_z,
\end{equation}
for traceless, Hermitian number preserving $k$ on $\mathcal{H}_{p-1} \otimes \mathcal{H}_p$.

We will work backwards starting from some $H( |\psi_0>, |\psi_1>, h)$ for $|\psi_0>, |\psi_1> \in S_{p n}$.
Since $|\psi_0>$ and $|\psi_1>$ have the same value of total $N$ on the region
$x \le p$, 
it follows that
a $U_0$ can be found in $\hat{G}_{p-1}$ such that
\begin{subequations}
\begin{eqnarray}
\label{u0psi0}
|\psi_2> & = & U_0 |\psi_0>, \\
\label{u0psi1}
|\psi_3> & = & U_0 |\psi_1>
\end{eqnarray}
\end{subequations}
are orthogonal vectors in $S_{p n}$ with equal total particle counts
on the region $p-1 \le x \le p$.
The particle count difference 
between $|\psi_0>$ and $|\psi_1>$ at point $p$ is at most
2, and equal and opposite to the difference between the corresponding
totals on the region $x \le p-1$. This compensating difference can be
moved by $U_0$ to the point $p -1$.
A $\hat{k}$ in Eq. (\ref{pminus1p}) can
then be found such that
\begin{subequations}
\begin{eqnarray}
\label{uone}
U_1 & = & \exp( i \hat{k}) \\
\label{u1psi2}
|\psi_4> & = & U_1 |\psi_2>, \\
\label{u1psi3}
|\psi_5> & = & U_1 |\psi_3>, \\
\label{phi4}
|\psi_4> & = & |\phi_4> \otimes |r>_p \\
\label{phi5}
|\psi_5> & = & |\phi_5> \otimes |r>_p,
\end{eqnarray}
\end{subequations}
for some $r \in \{-1, 0, 1, 2\}$
and $|\phi_4>$ and $|\phi_5>$ orthogonal vectors in $S_{p-1 m}$ with $m = n - |r|$.

It is then possible to find a $U_2$ in $\hat{G}_{p-1}$ such that
\begin{subequations}
\begin{eqnarray}
\label{u2psi4}
|\psi_6> & = & U_2 |\psi_4>, \\
\label{u2psi5}
|\psi_7> & = & U_2 |\psi_5>, \\
\label{psi6}
|\psi_6> & = & | \chi> \otimes |-1>_{p-1} \otimes |r>_p, \\
\label{psi7}
|\psi_7> & = & | \chi> \otimes |1>_{p-1} \otimes |r>_p, 
\end{eqnarray}
\end{subequations}
for a some $|\chi>$ in $S_{p-2 m-1}$.

Combining Eqs. (\ref{u0psi0}) - (\ref{psi7}), the induction hypothesis
implies unit determinant unitary operators $U_0, U_1, U_2$ such that
\begin{multline}
\label{combined}
U_2 U_1 U_0 H( |\psi_0>, |\psi_1>, h) U_0^\dagger U_1^\dagger U_2^\dagger = \\
|\chi>< \chi| \otimes \sum_{ij} |i>_{p-1}<j|_{p-1} h_{ij} \\
\otimes |r>_p<r|_p \otimes_{z > p} I_p,
\end{multline}
for $i$ and $j$ now summed over $-1$ and $1$ with the original index value 0 of $h_{ij}$ replaced by $-1$.

The expression on the right-hand side of Eq. (\ref{combined}) can then be obtained from a
commutator between an operator $\hat{k}$ from Eq. (\ref{pminus1p}) and
an operator $\hat{g} \in L_{p-1 m}$ for $m = n - |r|$.
For 2 x 2 traceless Hermitian $k_{ij}$, define
\begin{multline}
\label{kinpminus1p}
\hat{k} = \sum_{ij} |i>_{p-1}<j|_{p-1} k_{ij} \otimes \\
|r>_p <r|_p \otimes_{z \ne p-1,p} I_z,
\end{multline}
and for a 2 x 2 traceless Hermitian $g_{ij}$, define
\begin{multline}
\label{ginl}
\hat{g}  =  |\chi><\chi| \otimes \\ 
 \sum_{ij} |i>_{p-1}<j|_{p-1} g_{ij} \otimes_{z > p-1} I_z.
\end{multline}
Choose $k_{ij}$ and $g_{ij}$ so that their
commutator as 2 x 2 matrices fulfills
\begin{equation}
\label{hkg}
h = i [ k, g].
\end{equation}
Combining Eqs. (\ref{combined}), (\ref{kinpminus1p}), (\ref{ginl}) and (\ref{hkg}) then gives
\begin{equation}
\label{endresult}
H( |\psi_0>, |\psi_1>, h) = 
U_0^\dagger U_1^\dagger U_2^\dagger i[\hat{k}, \hat{g}] U_2 U_1 U_0,
\end{equation}
which completes the induction step and therefore the proof of Eq. (\ref{formofga}).


\begin{thebibliography}{9}

\bibitem{Everett} H. Everett, Rev. Mod. Phys. 29, 454 (1957).
\bibitem{DeWitt} B. DeWitt, Physics Today 23, 30 (1970).
\bibitem{Zeh} H. D. Zeh, Found. Phys. 1, 69 (1970).
\bibitem{Zurek} W. H. Zurek, Phys. Rev. D24, 1516 (1981); 
Phys. Rev. D26; 1862 (1982) Mod. Phys. 75, 715 (2003).
\bibitem{Wallace} D. Wallace, Studies in the History and Philosophy of Modern Physics 34, 87 (2003).
\bibitem{Riedel} C. Jess Riedel, Phys. Rev. Lett. 118, 120402 (2017).
\bibitem{Nielsen} M. A. Nielsen, arXiv:quant-ph/050207.
\bibitem{Stoica} Earlier suggestions of the possibility of obtaining the macroscopic world purely from 
unitary time evolution of a corresponding initial state appear in
L.S. Schulman, Phys. Lett. A 102(9):396-400, 1984 and
C. Stoica, Quanta 5, 19 (2016); arXiv:1607.02076[quant-phys].
The second of these proposes unitary time evolution as
a possible realization of a many-worlds branch and suggests
the result would be a kind of hidden variable theory. 
These proposal do not, however, show that such initial states
exit or how to obtain them and
do not include the other components of the present article. 
\bibitem{Page} D. N. Page, arXiv:1108.2709v2 [hep-th].
\bibitem{Schlosshauer} M. Schlosshauer, Rev. Mod. Phys. 76, 1267 (2004).
\bibitem{Kent} A. Kent, Found. Phys. 43, 421 (2012); Phys. Rev. A 90, 0121027 (2014); 
Phys. Rev. A 96, 062121 (2017).
\bibitem{Divincenzo} D. P. DiVincenzo, Phys. Rev. A 51, 1015(1995), arXiv:cond-mat/9407022.
\end{thebibliography}
\end{document}